\documentclass[aps,preprint,amsmath,amssymb,t1,11pt]{revtex4}
\usepackage{graphicx}
\usepackage{epstopdf}
\usepackage{xcolor}
\usepackage{slashed}
\usepackage{multirow}
\usepackage{booktabs} 
\usepackage{adjustbox}
\usepackage{rotating}

\begin{document}
\draft

\title{Constraints on Anomalous Quartic Gauge Couplings via Electroweak Production of $\gamma\gamma jj$ at Future Proton-Proton Colliders}

\author{A. Senol}
\email[]{senol_a@ibu.edu.tr}
\affiliation{Department of Physics, Bolu Abant Izzet Baysal University, 14280, Bolu, T\"{u}rkiye.}

\author{M. Tekin}
\email[]{wustafaatekin@gmail.com}
\affiliation{Department of Physics, Bolu Abant Izzet Baysal University, 14280, Bolu, T\"{u}rkiye.}
\author{B. S. Ozaltay}
\affiliation{Department of Physics, Bolu Abant Izzet Baysal University, 14280, Bolu, T\"{u}rkiye.}
\author{H. Denizli}
\email[]{denizli_h@ibu.edu.tr}
\affiliation{Department of Physics, Bolu Abant Izzet Baysal University, 14280, Bolu, T\"{u}rkiye.}

\date{\today}

\begin{abstract}

The investigation of quartic gauge couplings provides a crucial test of the Standard Model and serves as a potential window into new physics at higher energy scales. Within the framework of Effective Field Theory, deviations from the SM can be parameterized through dimension-8 operators. In this study, we analyze the process $pp \rightarrow \gamma\gamma jj$ at the High-Luminosity Large Hadron Collider (HL-LHC) and the Future Circular Collider in hadron mode (FCC-hh) to probe the sensitivity to anomalous quartic gauge couplings (aQGCs), particularly $f_{T8}/\Lambda^4$ and $f_{T9}/\Lambda^4$. Monte Carlo simulations of signal and relevant backgrounds are performed using MadGraph for event generation, Pythia for parton showering and hadronization, and Delphes for detector simulation. A multivariate analysis based on Boosted Decision Trees is employed to optimize the signal-to-background discrimination, incorporating a comprehensive set of kinematic and reconstructed variables of the final state particles. Additionally, we evaluate unitarity-violating effects associated with dimension-8 operators by imposing energy cutoffs on the di-photon invariant mass. The expected exclusion and discovery significances are computed, accounting for systematic uncertainties to ensure a realistic assessment of collider reach. Our findings indicate that the FCC-hh offers significantly improved sensitivity compared to the HL-LHC and current experimental results by ATLAS, reinforcing its potential for probing aQGCs. Notably, even under a 10\% systematic uncertainty, our projected limits for FCC-hh at 95\% confidence level surpass the current best constraints reported by the ATLAS collaboration, highlighting the enhanced discovery prospects at future high-energy colliders.

\end{abstract}


\maketitle

\section{Introduction}
The non-Abelian nature of gauge interactions in the electroweak sector of the Standard Model (SM) provides unique opportunities to probe the physics beyond the SM due to their sensitivity to SM Quartic Gauge Couplings (QGCs) and to possible anomalous QGCs (aQGCs). The non-abelian gauge nature of the SM predicts the existence of four quartic gauge boson vertices at the tree level $WWVV$ ($V=W^{\pm},\gamma,Z$) in combination where the total charge is neutral. Moreover, the five neutral quartic gauge boson vertices in the $V'V'V'V'$ ($V'=\gamma,Z$) structure do not exist in SM but they can be induced by BSM physics and allowed by the new processes at high energy scales. Therefore, the possible deviation of neutral QGCs ($V'V'V'V'$) from the SM predictions would directly point to the existence of new physics involving massive particles decoupling at energy scales much larger than the energies of the investigated electroweak (EW) sector. Anomalous QGCs are defined in the model-independent Effective Field Theory (EFT) approach and used to search for indirect signs of new physics in different production mechanisms at high energy lepton and hadron colliders. Among the various production mechanisms, vector boson scattering (VBS) plays a crucial role in probing the non-Abelian gauge structure of the electroweak (EW) interaction. In proton-proton collisions, VBS occurs when vector bosons ($V = W, Z, \gamma$) are emitted by incoming quarks and subsequently interact, leading to distinctive final-state signatures \cite{Green:2016trm, Jager:2017owh, Oh:2019wyq}.

The VBS processes involving purely EW interactions at leading order (LO) have two crucial features; two outgoing vector bosons and scattered quarks appearing as jets that are widely separated in pseudo-rapidity ($\eta$) and have a large di-jet mass ($m_{jj}$). In addition to these features, the minimum hadronic activity between these two jets leads to a unique topology that not only isolates VBS processes from other interactions, but also helps to reduce backgrounds arising from other processes. Therefore, the limits on aQGCs have been extensively reported in VBS searches in both the ATLAS and CMS Collaboration, as well as in phenomenological studies. In particular, these studies have been carried out in the same-sign $WW$ \cite{ATLAS:2014jzl,CMS:2017fhs,CMS:2020gfh,ATLAS:2023sua,Chaudhary:2019aim}, $W^{\pm}W^{\pm}$ \cite{ATLAS:2016nmw,ATLAS:2016snd,CMS:2019qfk,Fabbrichesi:2015hsa,Perez:2018kav,Yang:2021ukg,Jiang:2021ytz,Apyan:2022gis}, $WZ$ \cite{ATLAS:2016snd,CMS:2019qfk,CMS:2020gfh}, $W\gamma$ \cite{CMS:2020ypo,CMS:2022yrl, ATLAS:2024bho,Guo:2020lim}, $ZZ$ \cite{CMS:2017zmo,CMS:2019qfk,CMS:2020fqz,ATLAS:2023dkz,Semushin:2024uzn} and $Z\gamma$ \cite{CMS:2020ioi,CMS:2021gme,ATLAS:2022nru,ATLAS:2023fxh,Yang:2021pcf} channels associated with two jets. Machine learning algorithms have been used in some of these analyses due to their effectiveness in solving various tasks, such as particle identification, object reconstruction, and event classification. Since the most challenging and the common one of these tasks is the discrimination of a rare new-physics signal over the much larger amount of known Standard Model (SM) background processes, we employ the Boosted Decision Tree (BDT) method from the Multivariate Analysis Toolkit (TMVA) for the analysis of the $\gamma\gamma j j$ final state at future hadron-hadron collider to obtain the sensitivity on the dimension-8 anomalous quartic gauge couplings.

The largest scientific instrument Large Hadron Collider (LHC) will be upgraded by increasing its integrated luminosity (total collisions created) by a factor of ten beyond its original design value and operated for proton–proton collisions with a 14 TeV centre-of-mass energy. This new configuration, known as High Luminosity LHC (HL-LHC), is accompanied by upgrade projects of all LHC experiments and by the LHC Injector Upgrade Project (LIU) \cite{ZurbanoFernandez:2020cco}.
It will sustain and expand the discovery potential of post-LHC circular high-energy hadron colliders with the expectation of deepening our understanding of electroweak symmetry breaking, Higgs couplings and the origin of new physics. After the completion of the LHC and High-luminosity LHC physics programmes, energy-limit collider projects with the potential to search for new physics in a wide range of parameters are needed to precisely measure the Higgs self-coupling and to fully explore EWSB dynamics at the TeV scale. Therefore, the European Strategy Group's Update of the European Strategy for Particle Physics proposes to investigate the feasibility of a future circular collider. The Future Circular Collider (FCC) Study is one of the future project currently under consideration by CERN which comes to fore with its unique 100 km tunnel infrastructure and technology as well as the physics opportunities \cite{FCC:2018byv}. This project covers synergy and complementarity of different colliders options; a luminosity-frontier highest-energy lepton collider (FCC-ee ) with four proposed stages at different centre-of-mass energies \cite{FCC:2018evy} and  an energy-frontier hadron collider (FCC-hh) \cite{FCC:2018vvp}. The FCC-hh is designed to provide proton-proton collisions at the 100 TeV centre of mass energy with peak luminosity $5\times10^{34}$ cm$^{-2}$s$^{-1}$. 

In this paper, we present the effects of anomalous quartic gauge couplings on $ZZ\gamma\gamma$, $Z\gamma\gamma\gamma$ and $\gamma\gamma\gamma\gamma$ vertices via VBS process $pp\to \gamma\gamma jj$ as a potential of post-LHC circular high-energy hadron colliders; HL-LHC and FCC-hh. Section \ref{secII} introduces the Effective Field Theory (EFT) Lagrangian which defines the dimension-8 operators associated with the quartic gauge boson vertices. The methodology for event generation, event selection, and multivariate analysis is discussed in Section \ref{secIII}. In Section \ref{secIV}, we present the median expected significance for both discovery and exclusion, incorporating systematic effects. Finally, Section \ref{secV} concludes by comparing our results with the most stringent experimental limits.

\section{Theoretical background for anomalous Quartic Gauge Couplings}\label{secII}
 The Effective Field Theory (EFT) framework is used as a model-independent approach in which new particles cannot be directly produced. Instead, new physics arises through higher-dimensional operators involving Standard Model particles, as long as the energy scale of new physics is significantly higher than the centre-of-mass energy \cite{Buchmuller:1985jz,Hagiwara:1993ck}. The effective Lagrangian of the dimension-8 operators contributing to aQGCs can be written as \cite{Eboli:2006wa, Degrande:2013rea, Perez:2018kav}
\begin{eqnarray}\label{lag}
\mathcal{L}_{eff}=\mathcal{L}_{SM}+\sum_{j=0}^{1}\frac{f_{S,j}}{\Lambda^4}\mathcal{O}_{S,j}+\sum_{j=0}^{7}\frac{f_{M,j}}{\Lambda^4}\mathcal{O}_{M,j}+\sum_{j=0\atop \ j\neq 3}^{9}\frac{f_{T,j}}{\Lambda^4}\mathcal{O}_{T,j}
\end{eqnarray}
where the dimension-8 operators, $\mathcal O_{S,j}$ contains only covariant derivatives of the Higgs doublet, $\mathcal O_{M,j}$ contains two field strength tensors and two derivatives of the Higgs doublet, and $\mathcal O_{T,j}$ contains only field strength tensors. $f_{S,j}$, $f_{M,j}$ and $f_{T,j}$ represent the coefficients corresponding to the dimension-8 operators. $\Lambda$ is the scale of new physics. The explicit form of dimension-8 operators and corresponding quartic gauge boson vertices are given as in Table~\ref{operators}. 

\begin{figure}[h!]
\includegraphics[scale=0.6]{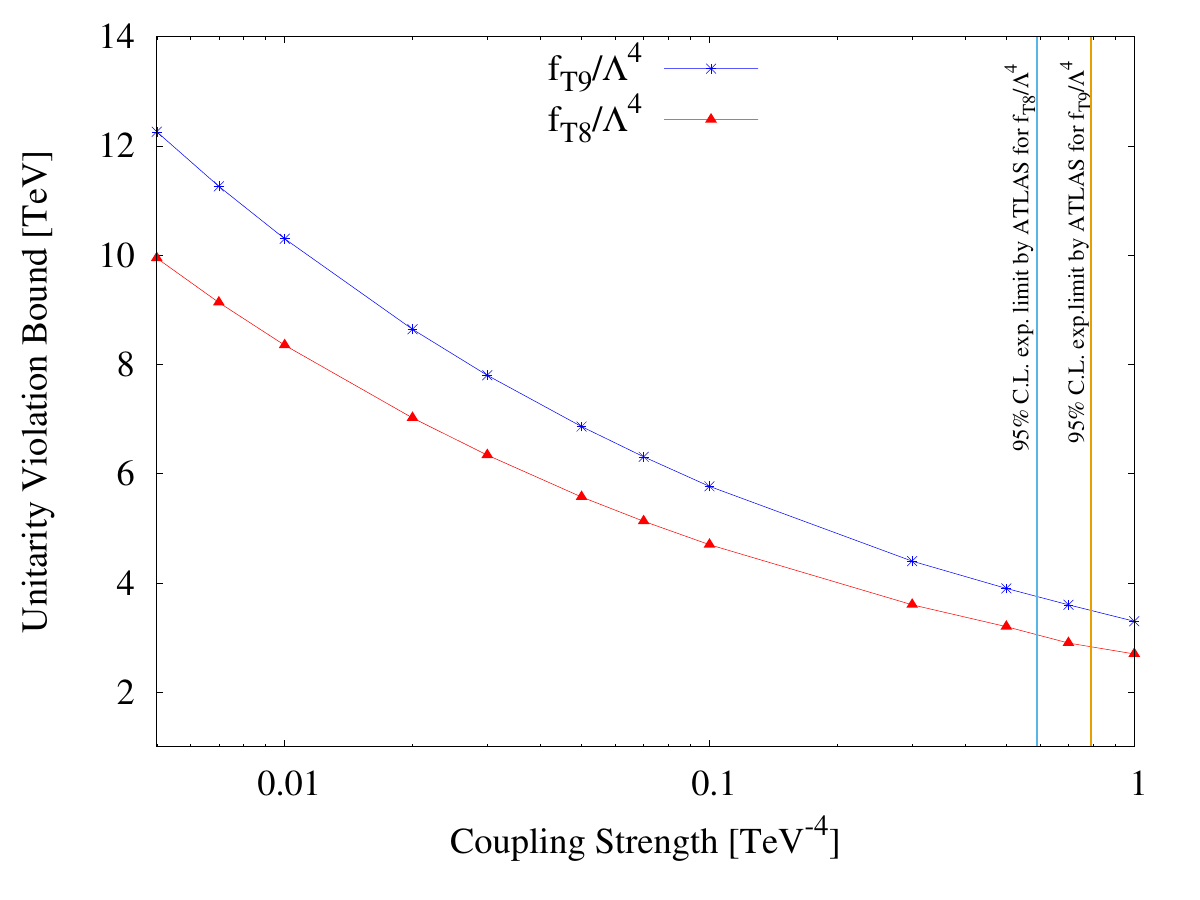}
\caption{The unitarity violation bound as a function of $f_{T8}/\Lambda^{4}$ and $f_{T9}/\Lambda^{4}$ aQGC via on-shell $VV\to VV$ scattering processes ($V = W / Z / \gamma$) calculated by combining channels with the same electric charge Q. Vertical lines represent one-dimensional experimental limits of $f_{T8}/\Lambda^{4}$ and $f_{T9}/\Lambda^{4}$ couplings where the unitarity is preserved  \cite{ATLAS:2022nru}.\label{unitarity}}
\end{figure}

\begin{table}[h]
\caption{Quartic gauge boson vertices modified by the explicit form of related dimension-8 operator \label{operators} }
\begin{ruledtabular}
\scriptsize{\begin{tabular}{lccccccccc}
Operators&\multicolumn{9}{c}{Vertices} \\
\hline
\multirow{1}{*}{$\mathcal{O}_{S0} = [(D_{\mu}\Phi)^{\dag}D_{\nu}\Phi]\times[(D^{\mu}\Phi)^{\dag}D^{\nu}\Phi]$}
 & $WWWW$ & $WWZZ$& $ZZZZ$ & & & & & & \\
\multirow{1}{*}{$\mathcal{O}_{S1} = [(D_{\mu}\Phi)^{\dag}D_{\nu}\Phi]\times[(D^{\mu}\Phi)^{\dag}D^{\nu}\Phi]$}   \\
\hline
\multirow{1}{*}{$\mathcal{O}_{M0} = \textrm{Tr}[\widehat{W}_{\mu\nu}\widehat{W}^{\mu\nu}]\times[(D_{\beta}\Phi)^{\dagger}D^{\beta}\Phi]$} \\
\multirow{1}{*}{$\mathcal{O}_{M1}=\textrm{Tr}[W_{\mu\nu}\widehat{W}^{\nu\beta}]\times[(D_{\beta}\Phi)^{\dagger}D^{\mu}\Phi]$}
  & $WWWW$ & $WWZZ$ & $ZZZZ$ & $WW \gamma Z$ & $WW \gamma \gamma$ & $ZZZ \gamma$ & $ZZ \gamma \gamma$ & & \\
\multirow{1}{*}{$\mathcal{O}_{M6} = [(D_{\mu}\Phi)^{\dagger}\widehat{W}_{\beta\nu}\widehat{W}^{\beta\nu} D^{\mu}\Phi]$} \\
\multirow{1}{*}{$\mathcal{O}_{M7} = [(D_{\mu}\Phi)^{\dagger}\widehat{W}_{\beta\nu}\widehat{W}^{\beta\mu} D^{\nu}\Phi]$} \\
\hline
\multirow{1}{*}{$\mathcal{O}_{M2} = [\widehat{B}_{\mu\nu}\widehat{B}^{\mu\nu}]\times[(D_{\beta}\Phi)^{\dagger}D^{\beta}\Phi]$} \\
\multirow{1}{*}{$\mathcal{O}_{M3}=[\widehat{B}_{\mu\nu}\widehat{B}^{\nu\beta}]\times[(D_{\beta}\Phi)^{\dagger}D^{\mu}\Phi]$} 
 & $WWZZ$ & $ZZZZ$ & $WW \gamma Z$ & $WW \gamma \gamma$ & $ZZZ \gamma$ & $ZZ \gamma \gamma$ & & \\
\multirow{1}{*}{$\mathcal{O}_{M4}=[(D_{\mu}\Phi)^{\dagger}\widehat{W}_{\beta\nu} D^{\mu}\Phi]\times \widehat{B}^{\beta\nu}$}\\
\multirow{1}{*}{$\mathcal{O}_{M5}=[(D_{\mu}\Phi)^{\dagger}\widehat{W}_{\beta\nu} D^{\nu}\Phi]\times \widehat{B}^{\beta\mu}+h.c$}\\
\hline
\multirow{1}{*}{$\mathcal{O}_{T0}=\textrm{Tr}[\widehat{W}_{\mu\nu}\widehat{W}^{\mu\nu}]\times \textrm{Tr}[\widehat{W}_{\alpha\beta}\widehat{W}^{\alpha\beta}]$}\\
\multirow{1}{*}{$\mathcal{O}_{T1}=\textrm{Tr}[\widehat{W}_{\alpha\nu}\widehat{W}^{\mu\beta}]\times \textrm{Tr}[\widehat{W}_{\mu\beta}\widehat{W}^{\alpha\nu}]$}
 & $WWZZ$ & $ZZZZ$ & $WW \gamma Z$ & $WW \gamma \gamma$ & $ZZZ \gamma$ & $ZZ \gamma \gamma$ & $Z\gamma\gamma\gamma$ & $\gamma\gamma\gamma\gamma$ & \\
\multirow{1}{*}{$\mathcal{O}_{T2}=\textrm{Tr}[\widehat{W}_{\alpha\mu}\widehat{W}^{\mu\beta}]\times \textrm{Tr}[\widehat{W}_{\beta\nu}\widehat{W}^{\nu\alpha}]$}\\
\hline
\multirow{1}{*}{$\mathcal{O}_{T5}=\textrm{Tr}[\widehat{W}_{\mu\nu}\widehat{W}^{\mu\nu}]\times \widehat{B}_{\alpha\beta}B^{\alpha\beta}$}\\
\multirow{1}{*}{$\mathcal{O}_{T6}=\textrm{Tr}[\widehat{W}_{\alpha\nu}\widehat{W}^{\mu\beta}]\times \widehat{B}_{\mu\beta}\widehat{B}^{\alpha\nu}$}
 & $ZZZZ$ & $WW \gamma Z$ & $WW \gamma \gamma$ & $ZZZ \gamma$ & $ZZ \gamma \gamma$ & $Z\gamma\gamma\gamma$ & $\gamma\gamma\gamma\gamma$ & \\
\multirow{1}{*}{$\mathcal{O}_{T7}=\textrm{Tr}[\widehat{W}_{\alpha\mu}\widehat{W}^{\mu\beta}]\times \widehat{B}_{\beta\nu}\widehat{B}^{\nu\alpha}$}\\
\hline
\multirow{1}{*}{$\mathcal{O}_{T8}=[\widehat{B}_{\mu\nu}\widehat{B}^{\mu\nu}\widehat{B}_{\alpha\beta}\widehat{B}^{\alpha\beta}]$}\\
\multirow{1}{*}{$\mathcal{O}_{T9}=[\widehat{B}_{\alpha\mu}\widehat{B}^{\mu\beta}\widehat{B}_{\beta\nu}\widehat{B}^{\nu\alpha}]$} & $ZZZZ$  & $ZZZ \gamma$ & $ZZ \gamma \gamma$ & $Z\gamma\gamma\gamma$ & $\gamma\gamma\gamma\gamma$ & 
\end{tabular}}
\end{ruledtabular}
\end{table}
The effective field theory remains valid below the scale of new physics, avoiding unitarity violation. However, high-dimensional operators with nonzero anomalous quartic gauge couplings (aQGCs) can result in a scattering amplitude that exceeds the unitarity limit at sufficiently high energies, known as the unitarity bound. For dimension-8 operators, the unitarity bound is determined using a dipole form factor:
\begin{eqnarray}
FF=\frac{1}{(1+\hat s/\Lambda_{FF}^2)^2}
\end{eqnarray}
Here, $\hat s$ represents the maximum centre-of-mass energy, and $\Lambda_{FF}$	is the form factor's energy scale. The maximum $\Lambda_{FF}$ is computed using VBFNLO 2.7.1 \cite{Arnold:2008rz}, which evaluates the zeroth partial wave of the amplitude from on-shell $VV\to VV$ scattering processes ($V = W / Z / \gamma$). The unitarity criterion requires the real part of the zeroth partial wave to remain below 0.5. VBFNLO combines all $VV$ channels with the same charge $Q$ and checks individual channels for unitarity. Fig.~\ref{unitarity} presents the unitarity violation (UV) bounds for the higher-dimensional operators considered in this study. Regions below the plotted lines for each coefficient indicate safe unitarity.

\begin{figure}[h!]
\includegraphics[scale=1.0]{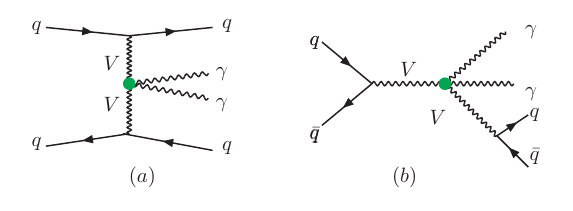}
\caption{Feynman diagrams illustrating the contribution of aQGC to (a) the EWK-VBS process and (b) the non-VBS process in $\gamma \gamma jj$ production at proton-proton collisions.\label{fd}}
\end{figure}
The production of $\gamma\gamma jj$ final states in proton-proton ($pp$) collisions occurs mainly through a combination of strong and electroweak interactions. In this work, we consider EWK $\gamma\gamma jj$ process containing at least two photon quartic gauge boson vertices at the tree level of order $\alpha_W^4$ (where $\alpha_W$ is the electroweak coupling constant). The $WW\gamma\gamma$, $Z\gamma\gamma\gamma$, $ZZ\gamma\gamma$ and $\gamma\gamma\gamma\gamma$ vertices contribute to both the $s$- and $t$-channel diagrams of the $pp\to\gamma\gamma jj$ process as illustrated in Fig.~\ref{fd} through the corresponding Feynman diagrams. There are also 4816 SM diagrams contributing to EWK $\gamma\gamma jj$ process based on initial and final quark combinations.

We focus on the two aQGCs, in particular $f_{T8}/\Lambda^{4}$ and $f_{T9}/\Lambda^{4}$, corresponding to neutral quartic gauge boson vertices ($ZZ\gamma\gamma$, $Z\gamma\gamma\gamma$, and $\gamma\gamma\gamma\gamma$) since the signal process $pp\to\gamma\gamma jj$ of interest in this work is more sensitive to neutral-weak-boson self-interactions arising due to the non-Abelian nature of the electroweak interaction. This sensitivity is due to anomalous quartic interactions which have different Lorentz structures than the corresponding interactions in the SM and can therefore distort the kinematic properties of the electroweak process, and their effects can be shown primarily in the cross sections.

\begin{figure}[h!]
\includegraphics[scale=0.27]{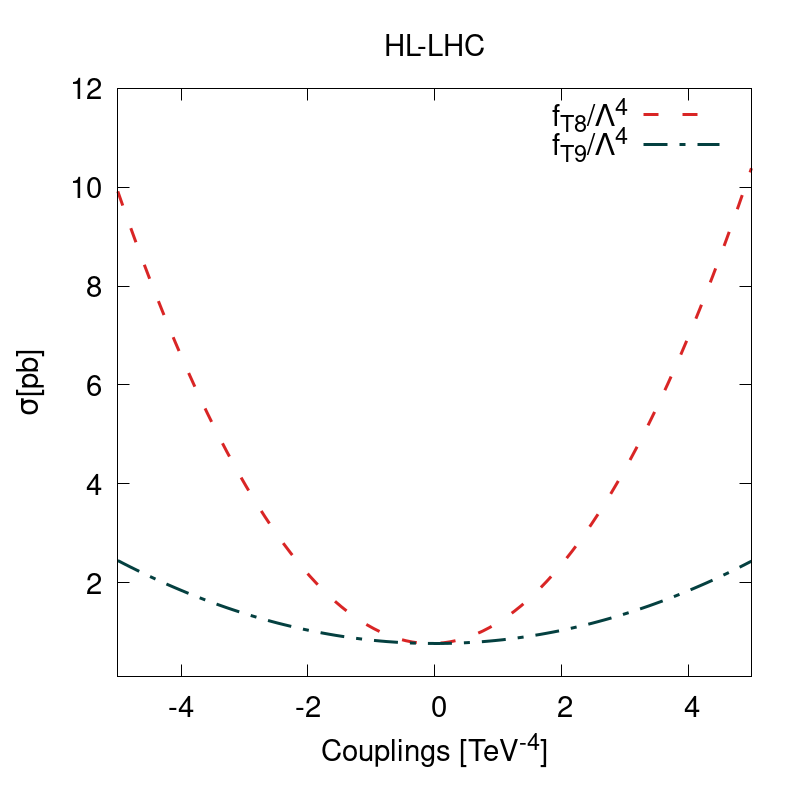}
\includegraphics[scale=0.27]{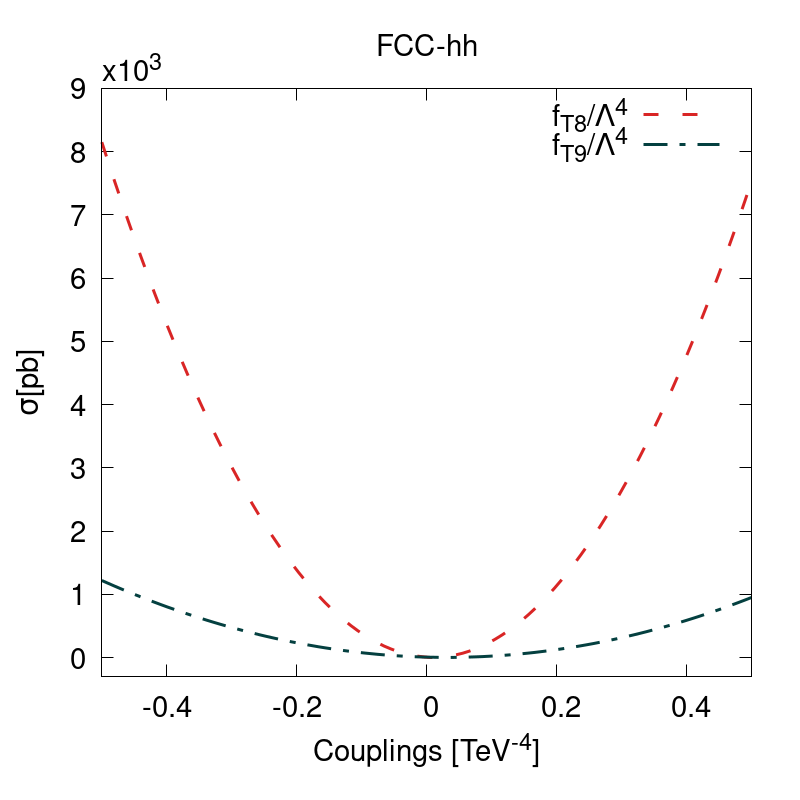}
\caption{The total cross sections of the EWK $pp \rightarrow \gamma\gamma jj$ process as a function of aQGC $f_{T8}/\Lambda^4$ and $f_{T9}/\Lambda^4$ at HL-LHC ($\sqrt{s}$=14 TeV) and FCC-hh ($\sqrt{s}$=100 TeV) \label{fig:xs_graphs}}
\end{figure}

For this purpose, signal events are generated in the {MadGraph5\_aMC$@$NLO v3\_1\_1} \cite{Alwall:2014hca} where the last part of effective Lagrangian in Eq.~(\ref{lag}) is implemented in the form of FEYNRULES \cite{Alloul:2013bka} package as the Universal FeynRules Output module (UFO) \cite{Degrande:2011ua} called  SM$\_$LT8$\_$LT9$\_$UFO (the anomalous $WW\gamma\gamma$ vertices is not included) \cite{lt8t9}. This module includes only $f_{T8}/\Lambda^{4}$ and $f_{T9}/\Lambda^{4}$ contributing to the SM Lagrangian. The signal cross sections are calculated at generator level including transverse momentum ($p_T^{\gamma} >$ 10 GeV, $p_T^{j} >$  20 GeV), pseudo-rapidity ( $|\eta^{j,\gamma}|< $ 2.5) of the photons and jets, and the minimum distance between each photon and jet  ($\Delta R(\gamma,j)>$ 0.4, $\Delta R(\gamma,\gamma)>$ 0.4 and $\Delta R(j,j)>$ 0.4) where $\Delta R(i,j)=[(\Delta\Phi_{i,j})^2 +(\Delta\eta_{i,j})^2]^{1/2}$ with $\Delta\Phi_{i,j}$ and $\Delta\eta_{i,j}$ are the azimuthal angle and the pseudo-rapidity difference between any two photons (two jets or photon and jet). Fig.~\ref{fig:xs_graphs} illustrates the dependence of the total cross section for the electroweak \( pp \to \gamma\gamma jj \) process on the anomalous quartic gauge couplings, \( f_{T8}/\Lambda^4 \) and \( f_{T9}/\Lambda^4 \), at two different collider scenarios: HL-LHC (\(\sqrt{s} = 14\) TeV) and FCC-hh (\(\sqrt{s} = 100\) TeV). In both cases, the cross section exhibits a quadratic dependence on the couplings, with a more pronounced increase for larger absolute values. Deviations from the SM cross section become apparent within the experimentally allowed range of \( f_{T8}/\Lambda^{4} \) and \( f_{T9}/\Lambda^{4} \) \cite{ATLAS:2022nru} which is at a magnitude comparable for HL-LHC but approximately an order of magnitude larger for FCC-hh. Moreover, the FCC-hh results in cross sections are approximately three orders of magnitude larger than those at HL-LHC in a given range. Additionally, the impact of \( f_{T8}/\Lambda^4 \) on the cross section is stronger compared to \( f_{T9}/\Lambda^4 \), as seen from the steeper slope of the red dashed curves. This indicates a greater sensitivity to \( f_{T8}/\Lambda^4 \) variations, which may provide enhanced opportunities for probing new physics effects at future high-energy hadron colliders.

The best available limits for $f_{T8}/\Lambda^{4}$ and $f_{T9}/\Lambda^{4}$ are obtained from a measurement of the fiducial cross section for the electroweak production $Z(\nu\bar\nu)\gamma jj$ by ATLAS collaboration in the region $p_T^{\gamma}>$ 150 GeV, with centre of mass energy of 13 TeV and integrated luminosity 139 fb$^{-1}$ are $[-5.2,5.2]\times 10^{-1}$ and $[-7.9,7.9]\times 10^{-1}$, respectively \cite{ATLAS:2022nru}. These limits preserve unitarity and were obtained by setting all aQGCs to zero except one. Moreover, the limits, when unitarity is not preserved, is also reported by ATLAS collaboration as $[-5.9,5.9]\times 10^{-2}$ and $[-1.3,1.3]\times 10^{-1}$ for $f_{T8}/\Lambda^{4}$ and $f_{T9}/\Lambda^{4}$ couplings, respectively. We expect to obtain better limits than the current experimental limits at both HL-LHC and FCC-hh colliders on the $f_{T8}/\Lambda^{4}$ and $f_{T9}/\Lambda^{4}$ aQGCs, which are more sensitive to this channel, by the detailed analysis of $\gamma\gamma jj$ production channel with machine learning technique, which will be detailed in the next section.

\section{Event Selection and details of Multivariate Analysis}\label{secIII}

We study the effects of anomalous quartic gauge couplings $f_{T8}/\Lambda^{4}$ and $f_{T9}/\Lambda^{4}$ on the $ZZ\gamma\gamma$, $Z\gamma\gamma\gamma$ and $\gamma\gamma\gamma\gamma$ vertices through the signal $pp\to \gamma\gamma jj$ and relevant background processes in details by including parton showering and the response of detector effects at the post-LHC circular high-energy hadron colliders (HL-LHC and FCC-hh). The generation of EWK signals and backgrounds at the lowest tree level is of the order of $\alpha_W^4$, while in QCD-induced background generation mechanisms it is of the order of $\alpha_W^2\alpha_s^2$, where $\alpha_W^4$ and $\alpha_s^2$ are the respective electroweak and strong coupling constants. The EWK $\gamma\gamma jj$ process including  non-zero aQGC parameters is considered as a signal including SM contribution as well as interference between effective couplings and SM contributions (($\gamma\gamma jj$)EWK). Since the  signal process is characterized by two photons and two jets in the final state, the relevant backgrounds shown in the Table ~\ref{event_gen} are considered for further  analysis. Among these backgrounds, ($\gamma\gamma jj$)EWK and ($\gamma\gamma jj$)QCD are the main backgrounds with the same final state as the signal process. 

\begin{table}[h]
\caption{Summary of signal and relevant background processes generated in {\sc MadGraph5\_aMC@NLO} and used in the analysis \label{event_gen}.}
\begin{ruledtabular}
\begin{tabular}{l c l l}
&Physical Process & Short name & Interaction order  \\ \hline
Signal& $pp\to\gamma\gamma jj$ (non-zero aQCs, interferences) & ($\gamma\gamma jj$)EWK & $\alpha_W^4$  \\ \hline
&$pp\to\gamma\gamma jj$ & ($\gamma\gamma jj$)EWK & $\alpha_W^4$\\
&$pp\to\gamma\gamma jj$ &($\gamma\gamma jj$)EWK & $\alpha_W^2\alpha_s^2$\\
Backgrounds & $pp\to W\gamma jj$ &$(W\gamma jj)$EWK &$\alpha_W^4$\\
&$pp\to W\gamma jj$ &$(W\gamma jj)$QCD &$\alpha_W^2\alpha_s^2$\\
&$pp\to Z\gamma jj$ &$(Z\gamma jj)$EWK &$\alpha_W^4$\\
&$pp\to Z\gamma jj$ &$(Z\gamma jj)$QCD &$\alpha_W^2\alpha_s^2$
\end{tabular}
\end{ruledtabular}
\end{table}

To conduct a detailed analysis of signal and background processes, approximately 500k events are generated using {\sc MadGraph5\_aMC@NLO} with the NNPDF23LO PDF set \cite{Ball:2012cx}. This is done for each background process as well as signal, corresponding to different values of the $f_{T8}/\Lambda^{4}$ and $f_{T9}/\Lambda^{4}$ couplings. Parton level events are then passed through to {\sc Pythia 8.20} \cite{Sjostrand:2014zea} for parton showering and hadronization. Detector responses which are parametrized in the form of resolution functions and efficiencies are included by {\sc Delphes 3.4.2} \cite{deFavereau:2013fsa} software with the corresponding Delphes cards, namely \verb|delphes_card_HLLHC.tcl| and \verb|FCC-hh.tcl|. Jets are reconstructed by using clustered energy deposits with {\sc FastJet 3.3.2} \cite{Cacciari:2011ma} using anti-kt algorithm \cite{Cacciari:2008gp} where a cone radius is set as $\Delta R$ = 0.4 and $p_T^j>$ 25 GeV.

\begin{figure}[h!]
\includegraphics[scale=0.12]{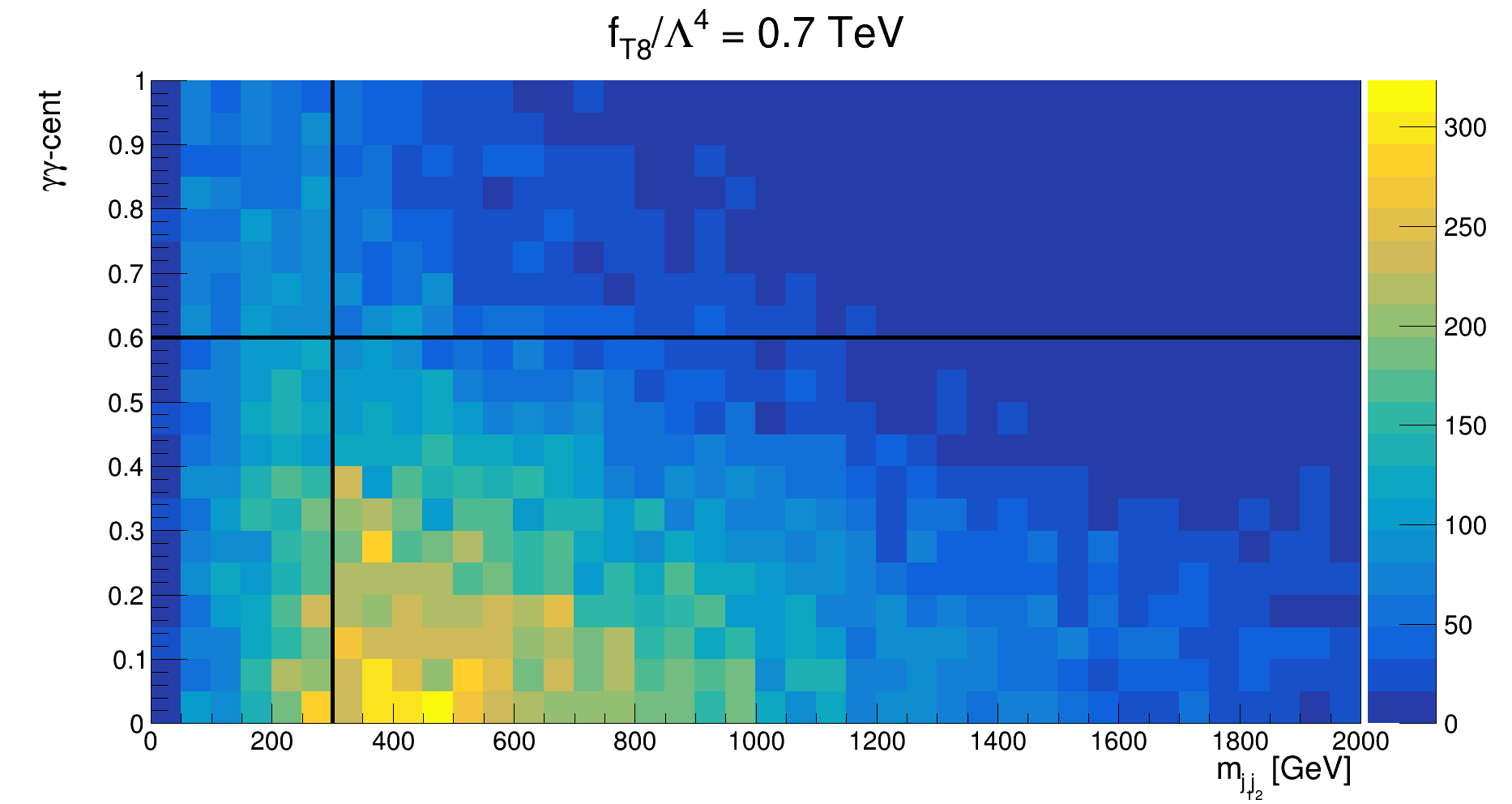}
\includegraphics[scale=0.12]{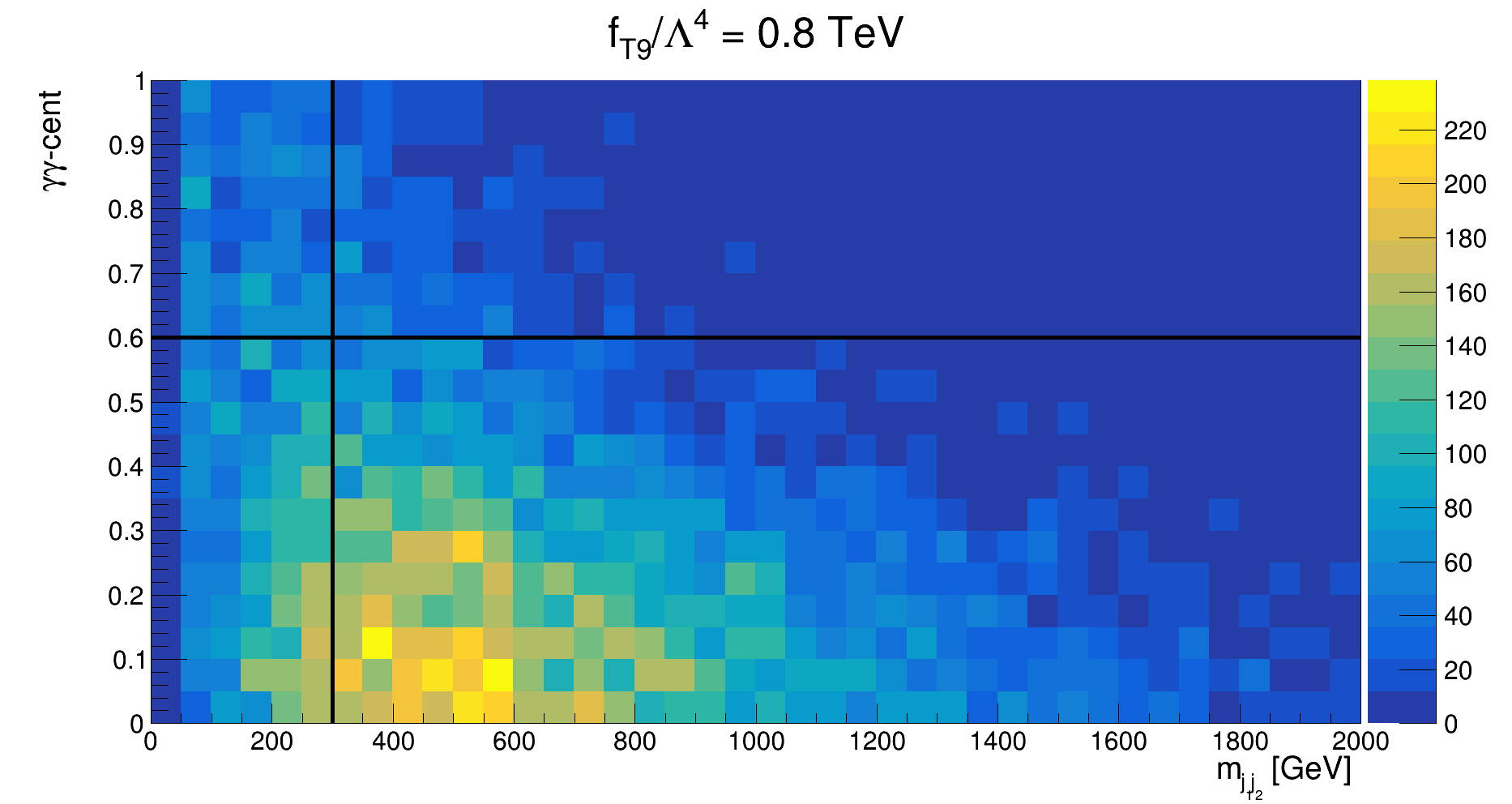}
\includegraphics[scale=0.12]{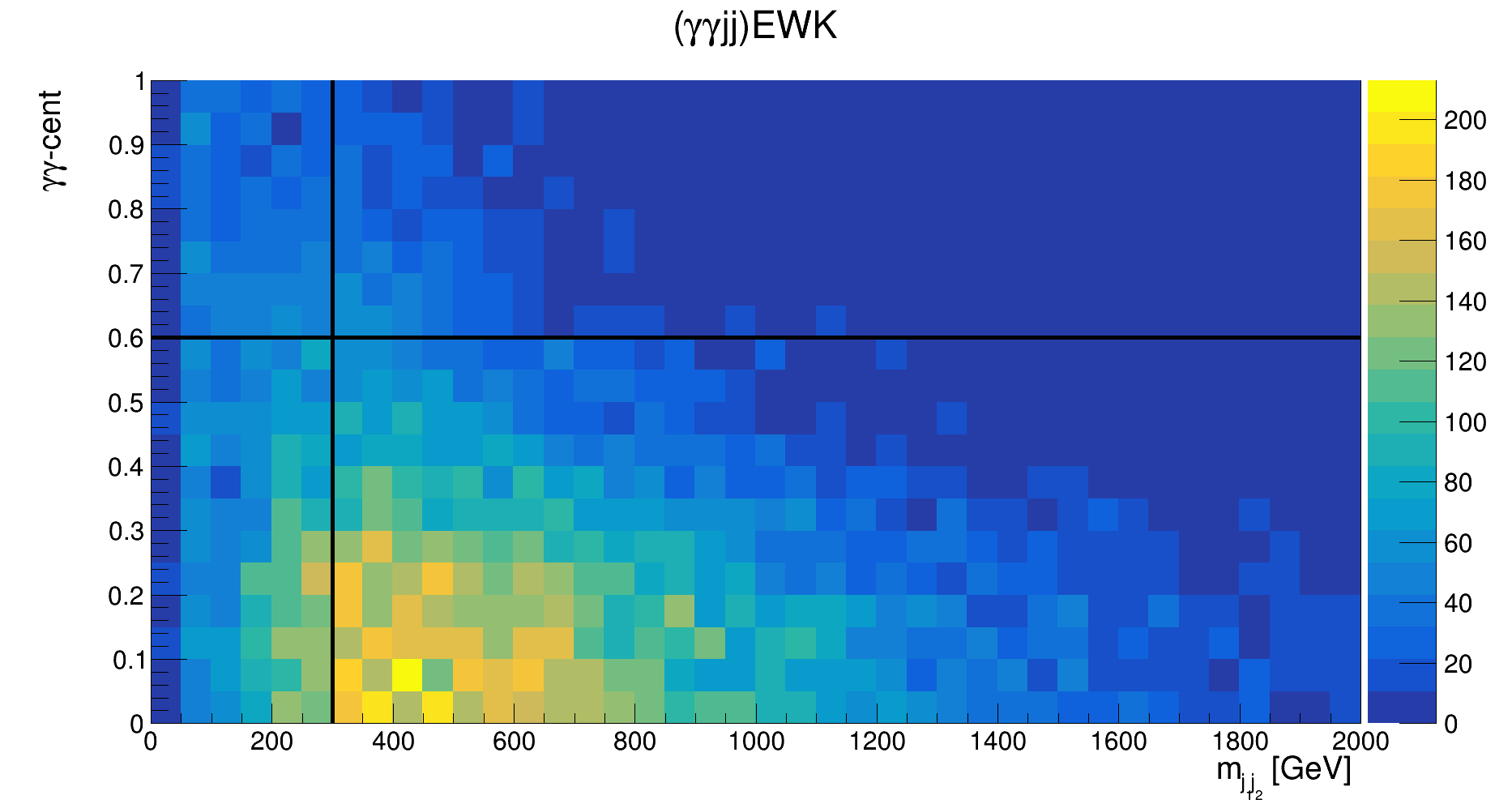}
\includegraphics[scale=0.12]{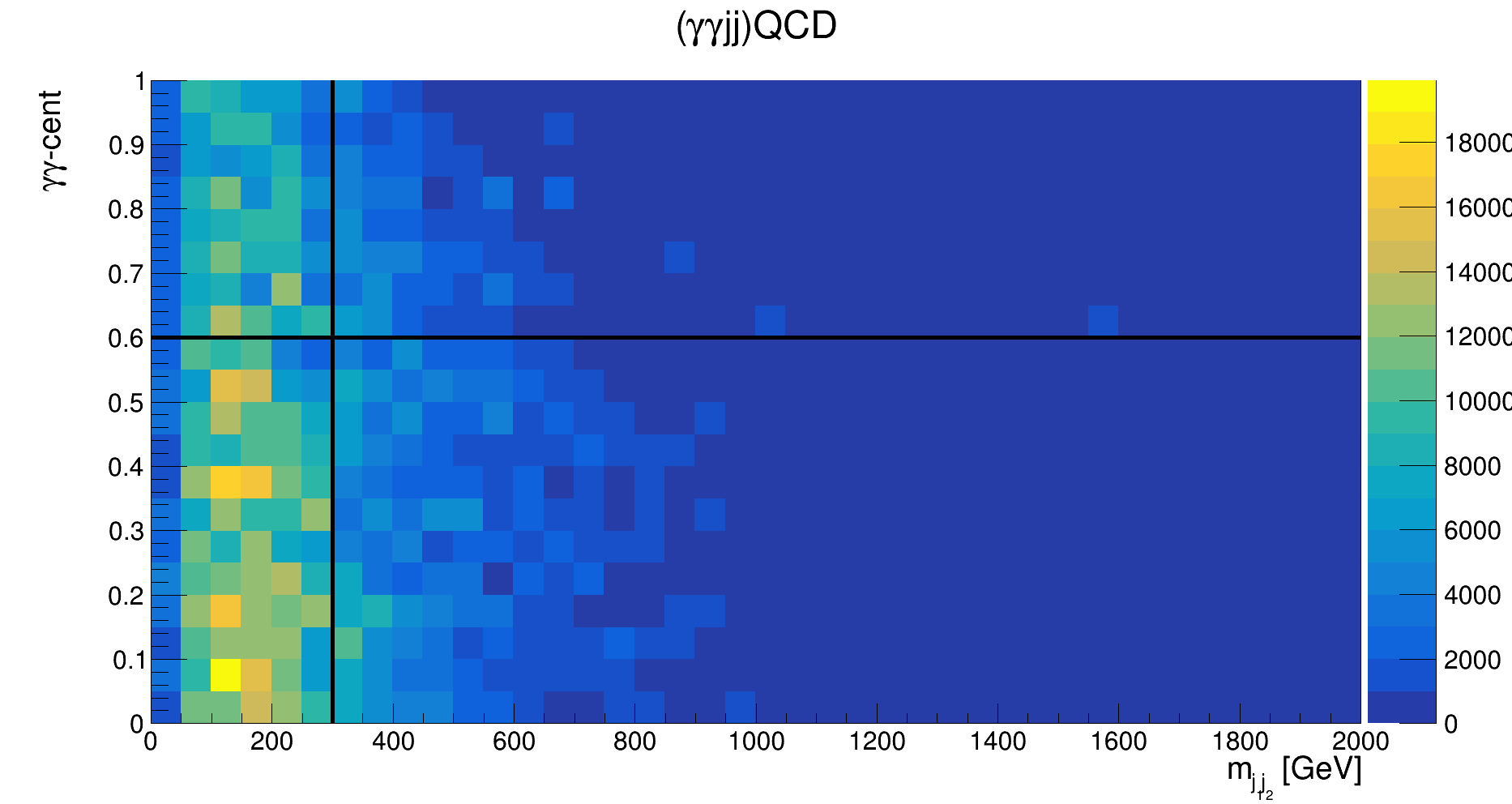}
\includegraphics[scale=0.12]{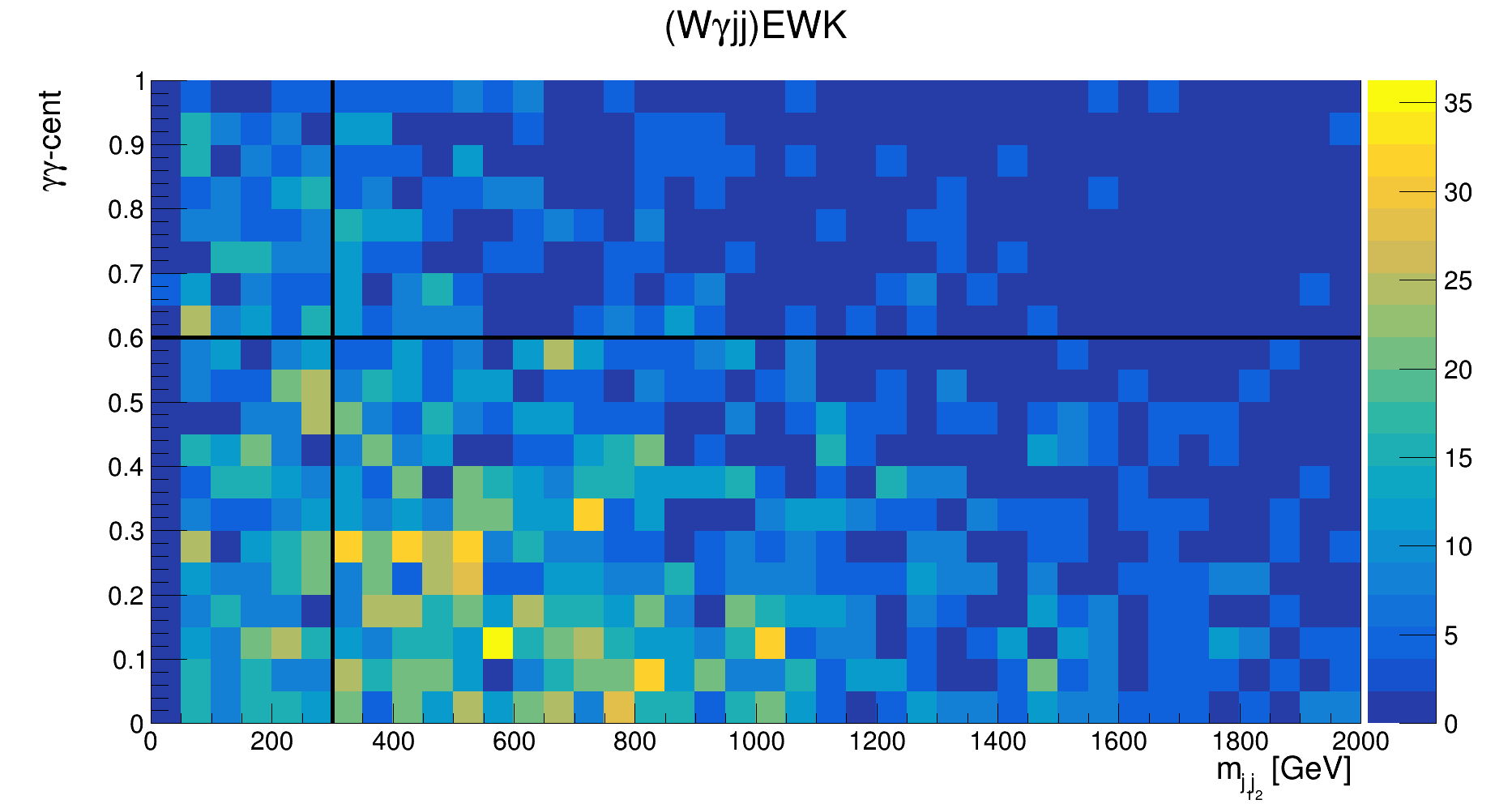}
\includegraphics[scale=0.12]{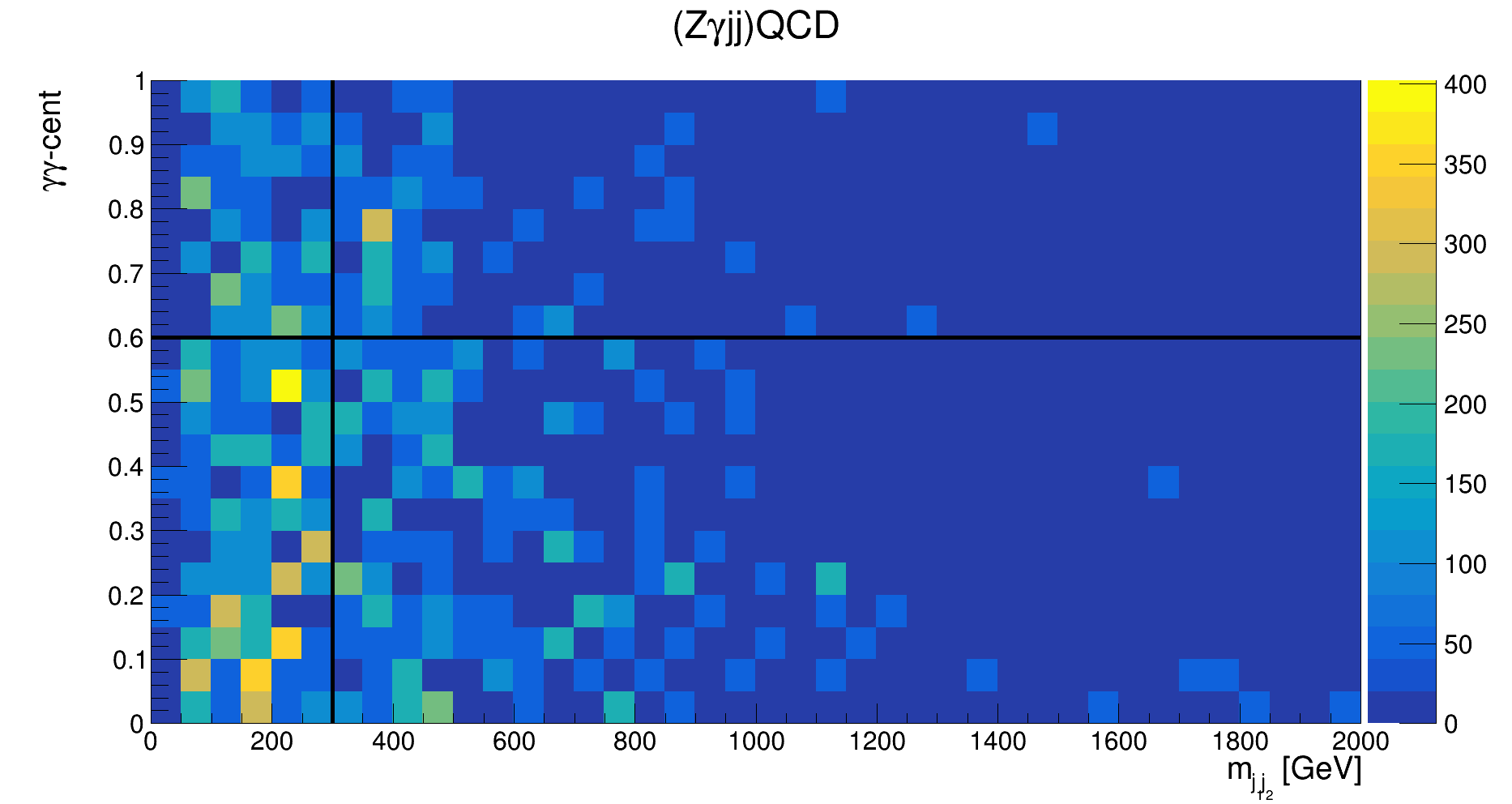}
\includegraphics[scale=0.12]{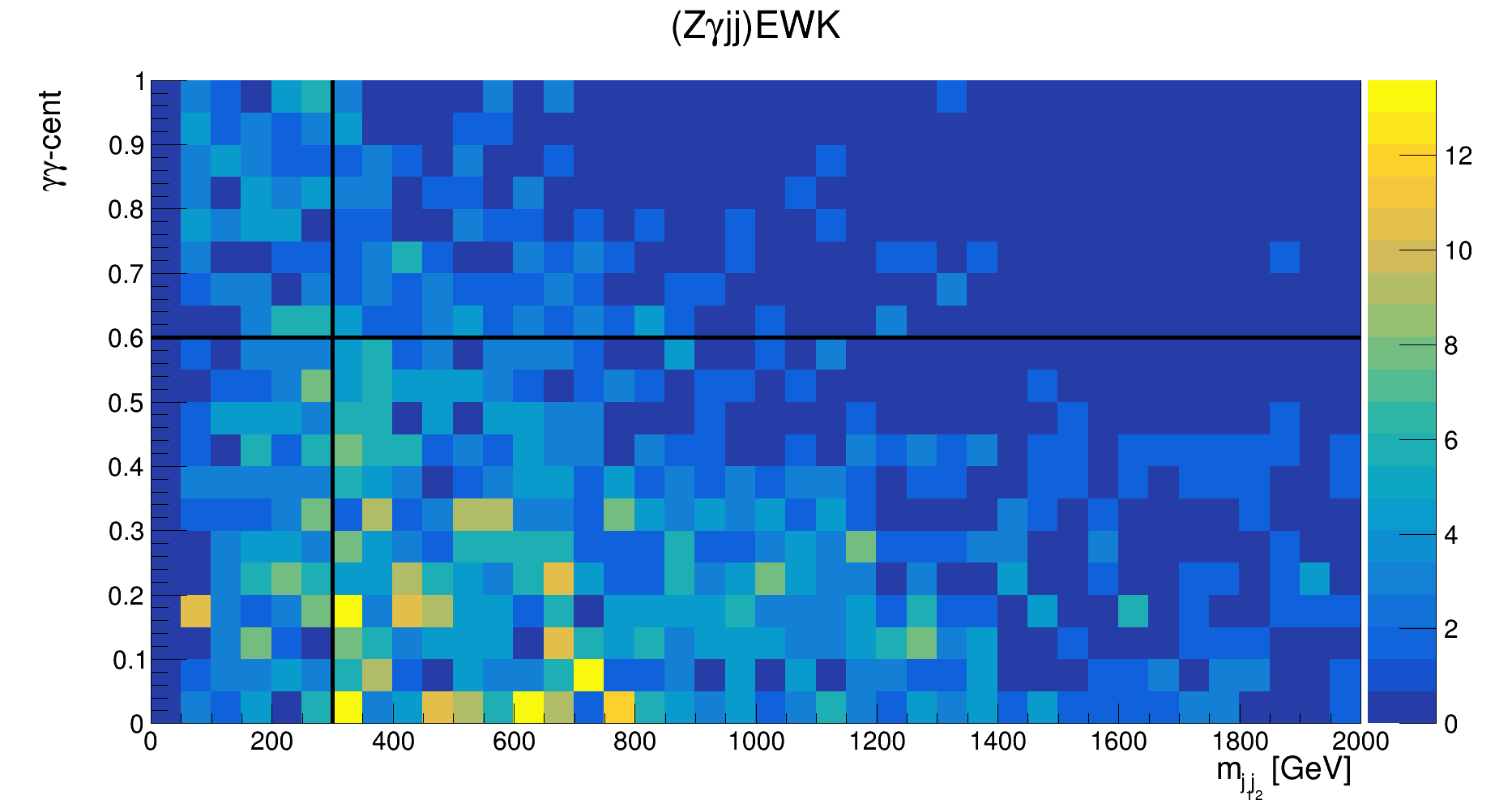}
\includegraphics[scale=0.12]{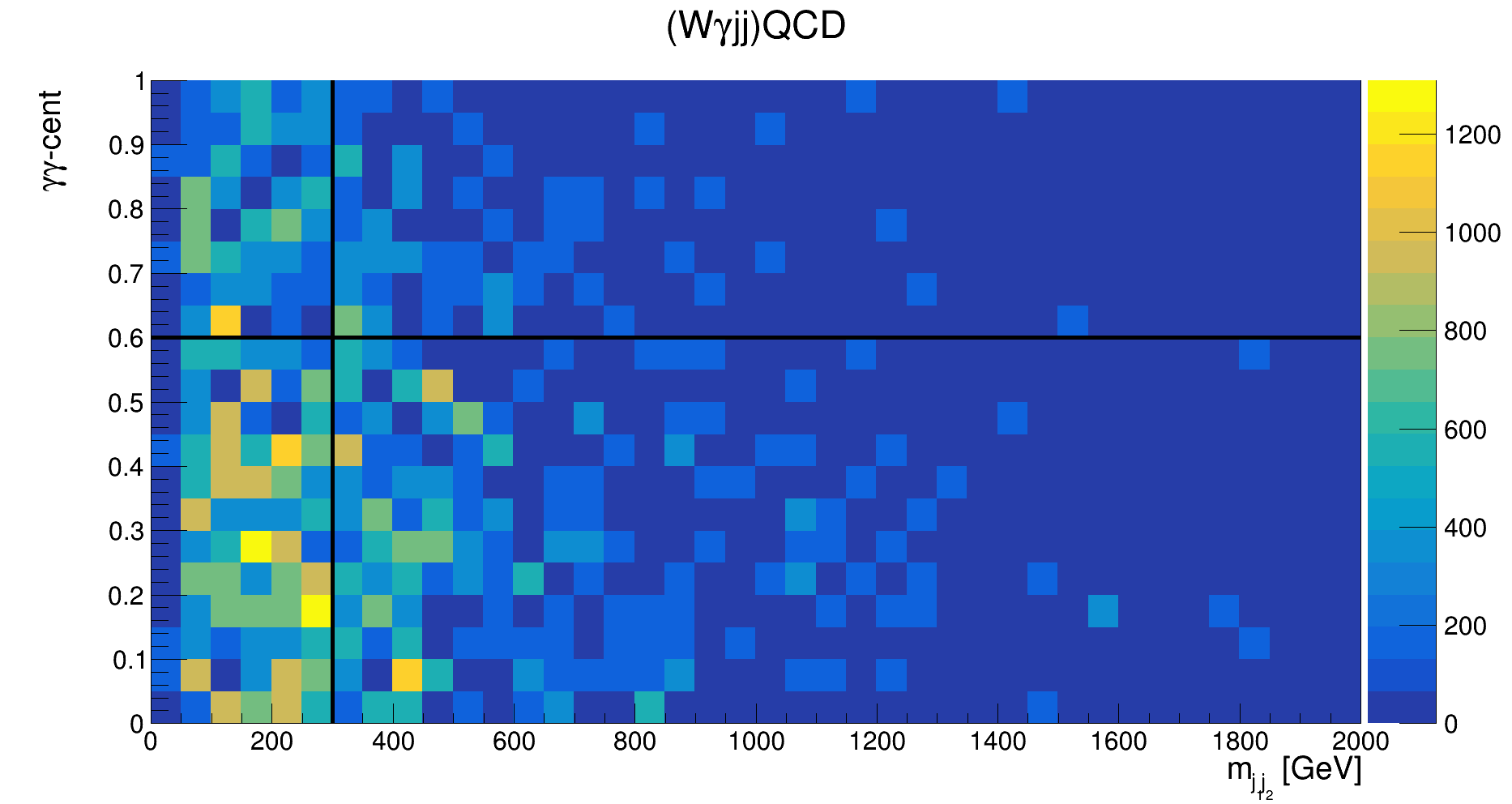}

\caption{The centrality of the two-photon system ($\gamma \gamma -cent$) versus invariant mass of two-jet system ($m_{j_1 j_2}$) for the signal ($f_{T8}/\Lambda^{4}$ = 0.7 $\text{TeV}^{-4}$, $f_{T9}/\Lambda^{4}$ = 0.8 $\text{TeV}^{-4}$) and relevant background processes at HL-LHC. \label{mjj_HL_LHC}}
\end{figure}

The event pre-selection of the analysis is initiated with the necessity of at least two photons ($N_{\gamma} > 1$) and two jets ($N_j > 1$) being present in the final state. A lepton veto ($N_l = 0$) is applied to suppress the contributions coming mainly from the SM $W \gamma jj$ and $Z \gamma jj$ background processes, which leads to the rejection of events containing any electron or muon present in the final state. 
\begin{table}[h!]
    \caption{Event selection and applied kinematic cuts used in the analysis for the HL-LHC (FCC-hh) colliders before the multivariate analysis \label{tab_event_selection}.} 
    \begin{ruledtabular}
    \begin{tabular}{l l }
        \textbf{Cuts} &  Parameters \\
        \hline
        \hline
        Event Pre-selection &  $N_j> 1$ \&  $N_{\gamma} > 1$  \& $N_l=0$  \\ 
        \hline
       & $p_T^{\gamma_1} > 150$ GeV $p_T^{\gamma_2} > 10$ GeV  \\
         Kinematic Cuts  & $p_T^{j_1} > 20$ GeV $p_T^{j_2} > 10$ GeV    \\
         &\( |\eta^{\gamma_1,\gamma_2,j_1,j_2}| \leq 2.5\)  \\\hline
     Reconstructed Cuts     & $m_{j_1 j_2} > 300$ GeV \\
         & $\gamma\gamma-cent<$ 0.6 (0.4) 
    \end{tabular}
    \end{ruledtabular}
\end{table}
By considering typical requirements of the experimental detectors and normalized distributions of final state jets and photons, we impose minimal cuts on the transverse momentum and rapidity of the final-state jets and photons given as kinematic cuts in Table~\ref{tab_event_selection} for each collider options. In addition to having a large  di-jet invariant mass ($m_{j_{1}j_{2}}$) as being the characteristic signatures of VBS processes, the centrality of the two-photon system ($\gamma \gamma -cent$) relative to the two highest-$p_{T}$ jets, the characteristic of the signal process, is defined as follows: 
\begin{eqnarray}
    \gamma\gamma-cent = \left |\frac{y(\gamma_1\gamma_2) - 0.5[y(j_1) + y(j_2)]}{y(j_1)- y(j_2)}\right |
\end{eqnarray}
where $y(j_1)(y(j_2))$ is rapidity of the leading (sub-leading) jet and  $y(\gamma_1\gamma_2)$ is the rapidity of the $\gamma_1\gamma_2$ system. In Fig.~\ref{mjj_HL_LHC}, we show the two dimensional distribution of two photon-system $\gamma \gamma - cent$ versus invariant mass of two-jet system $m_{j_{1} j_{2}}$ for the signal $f_{T8}/\Lambda^{4}$ = 0.7 $\text{TeV}^{-4}$, $f_{T9}/\Lambda^{4}$ = 0.8 $\text{TeV}^{-4}$ and relevant background processes at HL-LHC. Therefore, we define a signal region (represent with solid black line) as in the Fig.~\ref{mjj_HL_LHC}, $m_{j_1j _2}$ greater than 300 GeV and the centrality of the two-photon system ($\gamma \gamma -cent$) relative to the two highest-$p_{T}$ jets less than 0.6 for HL-LHC to suppresses the contributions coming from low-energy region in QCD background processes. A similar distribution is obtained for FCC-hh and the signal region is selected as $m_{j_1j_2}>300$ GeV and $\gamma \gamma -cent < 0.4$ in the analysis. 

\begin{figure}[h!]
\includegraphics[scale=0.407]{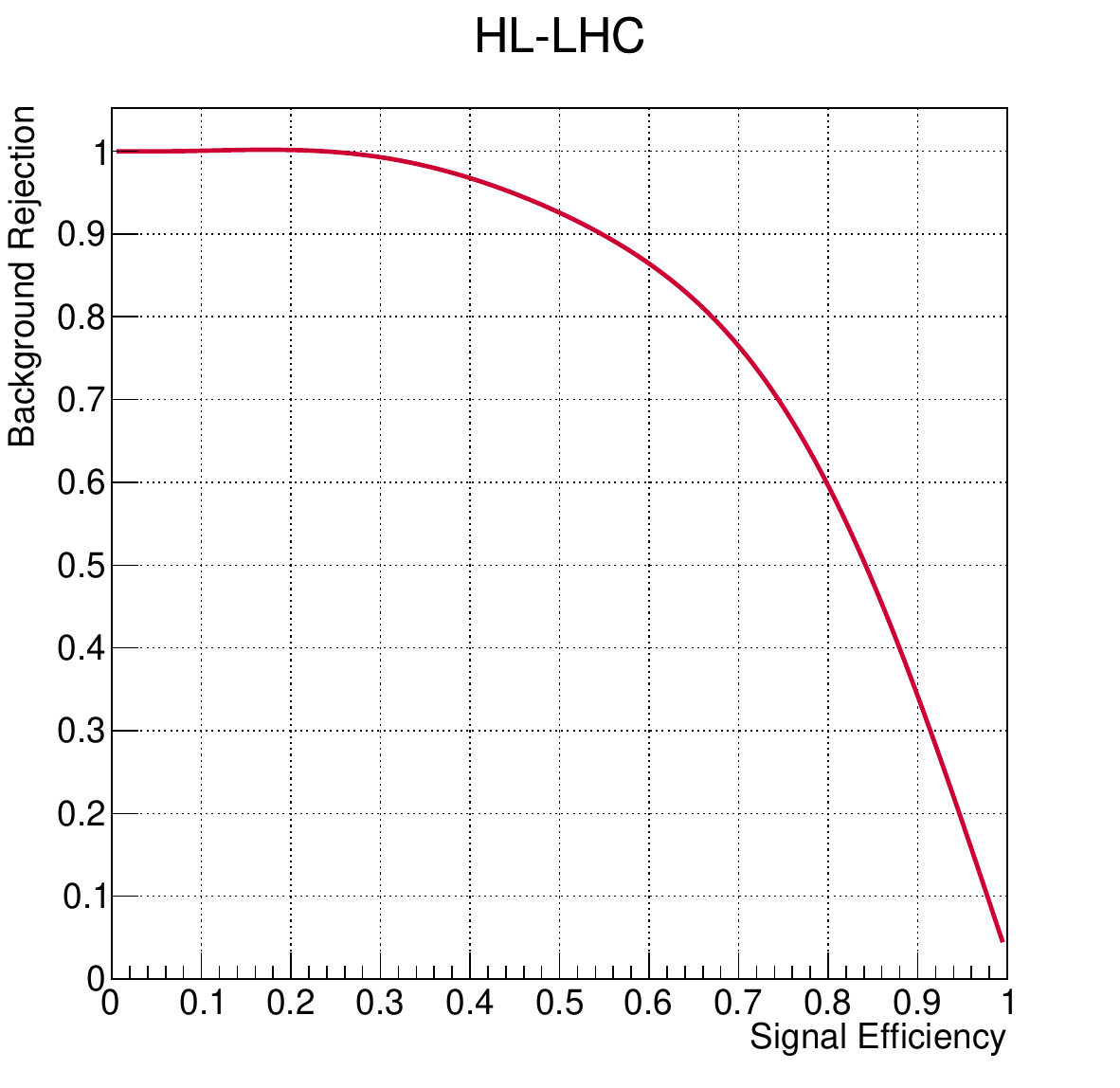}
\includegraphics[scale=0.407]{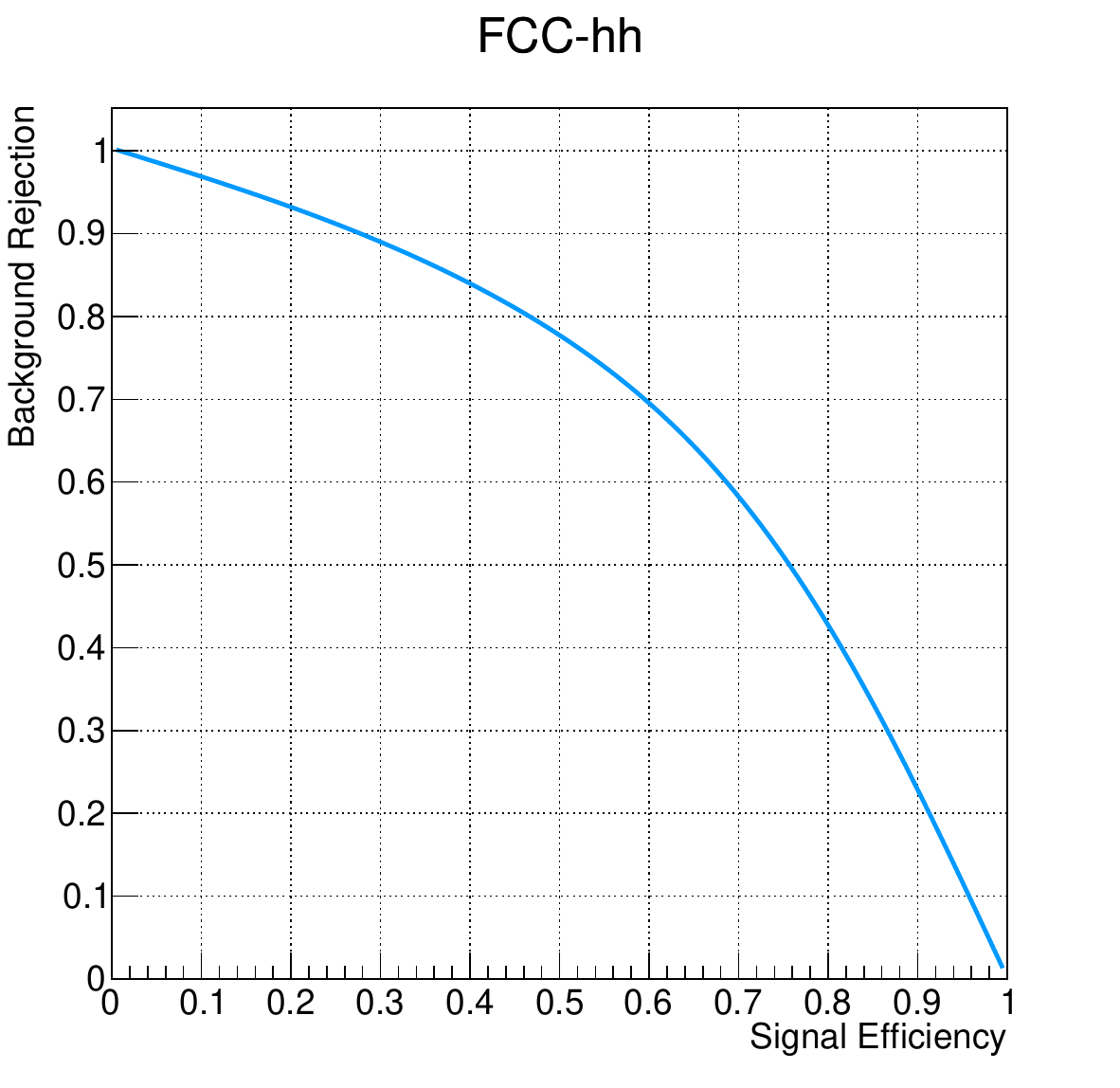}
\includegraphics[scale=0.407]{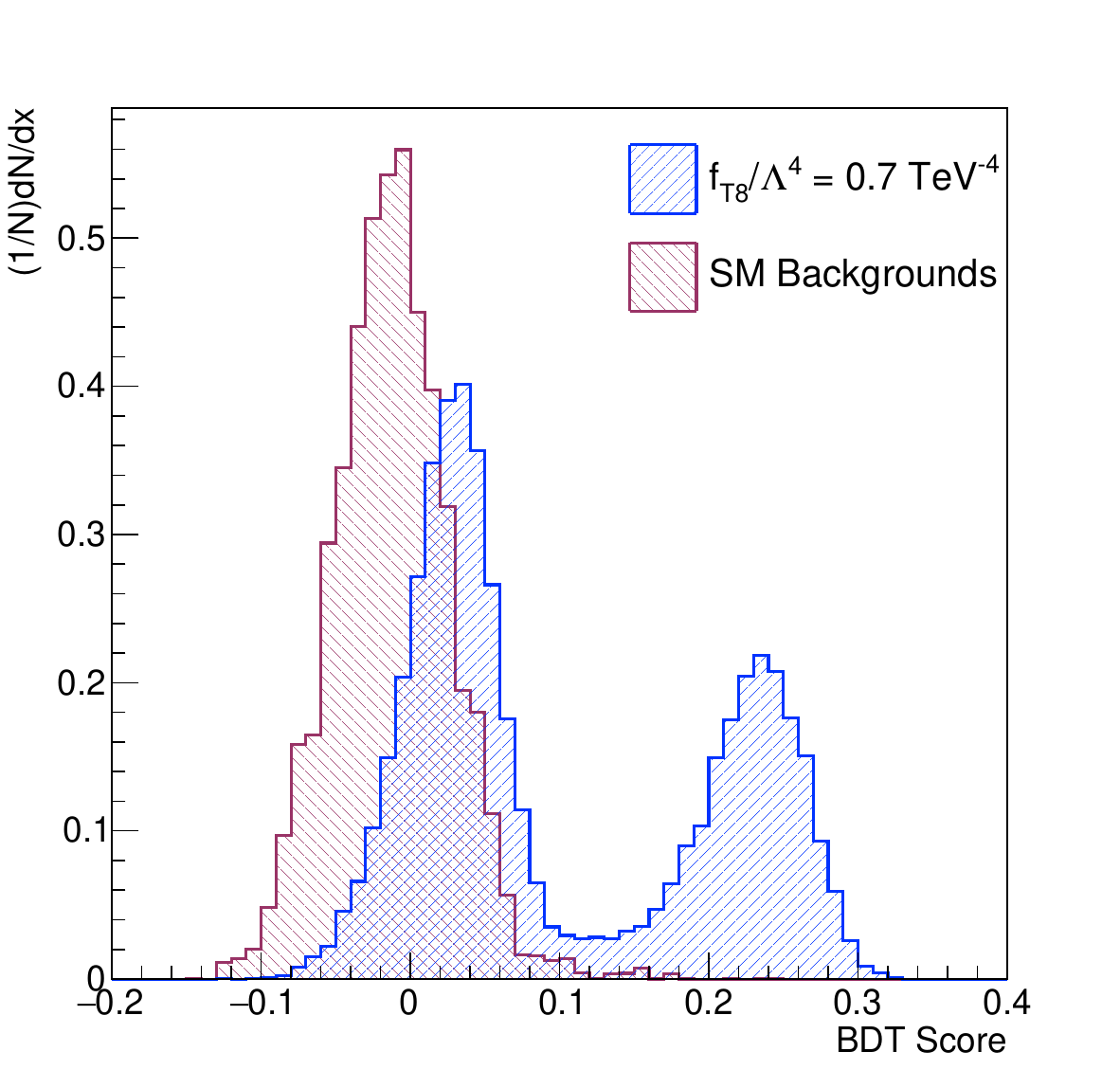}
\includegraphics[scale=0.407]{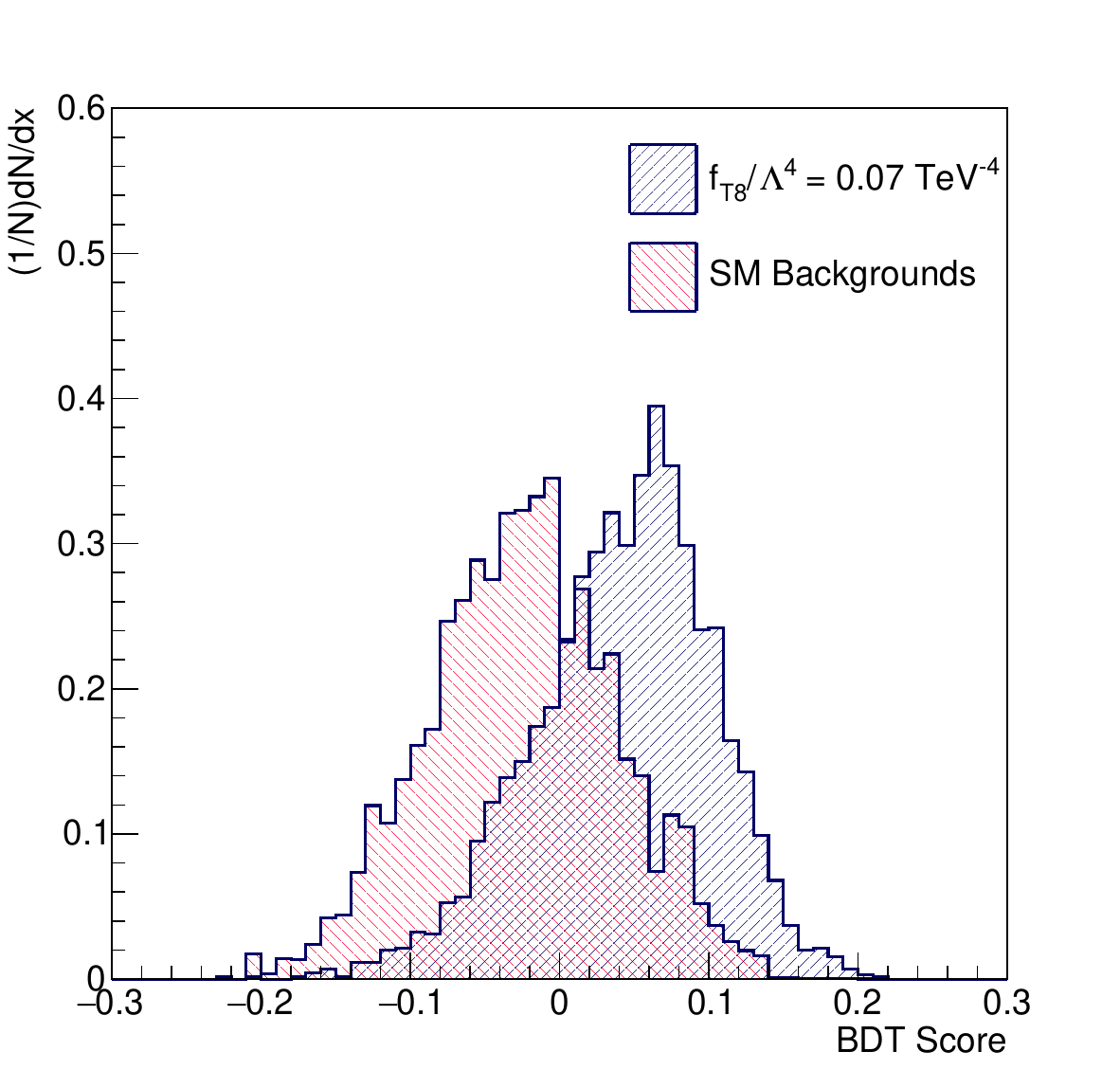}
\caption{The first row presents the Receiver Operating Characteristic (ROC) curve of the Boosted Decision Tree (BDT), while the second row illustrates the distribution of the BDT response for the signal (corresponding to \( f_{T8}/\Lambda^{4} = 0.7 \) TeV\(^{-4}\) and \( f_{T8}/\Lambda^{4} = 0.07 \) TeV\(^{-4}\)) along with all relevant background processes, analyzed for both the HL-LHC and FCC-hh.
 \label{BDT_output}}
\end{figure}

In our multivariate analysis, we employ the BDT machine learning technique with the following hyperparameters: 850 trees, a maximum depth of 3, a minimum of 2.5 events per terminal leaf, 20 iterations for optimal split identification, and a learning rate of 0.5. The training process utilizes adaptive boosting (AdaBoost), where signal events misclassified into background nodes are reweighted to improve performance. Since the choice of input features affects the discriminating performance of the BDT classifier, we include 20 kinematical and reconstructed variables from the leading  and sub-leading photons ($\gamma_1$, $\gamma_2$) and jets ($j_1$, $j_2$), respectively. Detailed description of the BDT variables are given in Table~\ref{bdtlist}. We combine data on this variable list for signal events as well as the background events in one data file. All the events are required to have cuts summarized in Table~\ref{tab_event_selection}. The background events are combined after properly weighted according to their cross-sections. To avoid unitarity violation in this analysis, the clipping method is employed. This approach \cite{ATLAS:2022nru,CMS:2020gfh} suppresses any effective field theory (EFT) contributions beyond a defined energy scale, $\Lambda_{FF}$. The value of $\Lambda_{FF}$ is determined for each aQGC parameter using the VBFNLO program, as discussed in the previous section. The corresponding clipping energy threshold for each aQGC parameter is then applied to the invariant mass of the two-photon system. To train the BDT classifier, we divide the data into two equal parts: 50\% is allocated for the training set, and the remaining 50\% is reserved for the test set. BDT response plots are shown in Fig.~\ref{BDT_output} for the signal and total background processes under consideration. 

\begin{table}[h!]
    \centering
    \caption{
    The input variable list of the BDT: the  kinematical and reconstructed variables from the leading and sub-leading photons ($\gamma_1$,$\gamma_2$) and jets ($j_1$,$j_2$).\label{bdtlist}} 
    \vspace{0.5em}
    \begin{tabular}{ll}
        \toprule
        \textbf{Variable} & \textbf{Definition} \\
        \hline
        \hline
        $p_T^{\gamma_1}$ & Transverse momentum of the leading photon \\
        $p_T^{\gamma_2}$ & Transverse momentum of the second-highest-$p_T$ photon \\
        $p_T^{j_1}$ & Transverse momentum of the leading jet \\
        $p_T^{j_2}$ & Transverse momentum of the second-highest-$p_T$ jet \\
        $\eta^{\gamma_1}$ & Pseudo-rapidity of the leading photon \\
        $\eta^{\gamma_2}$ & Pseudo-rapidity of the second-highest-$p_T$ photon \\
        $\eta^{j_1}$ & Pseudo-rapidity of the leading jet \\
        $\eta^{j_2}$ & Pseudo-rapidity of the second-highest-$p_T$ jet \\
        $\Delta R(\gamma_1, \gamma_2)$ & Distance between leading photon and sub-leading photon in $\eta$-$\phi$ plane \\
        $\Delta R(\gamma_1, j_1)$ & Distance between leading photon and leading jet in $\eta$-$\phi$ plane \\
        $\Delta R(\gamma_1, j_2)$ & Distance between leading photon and sub-leading jet in $\eta$-$\phi$ plane \\
        $\Delta R(\gamma_2 , j_1)$ & Distance between sub-leading photon and leading jet in $\eta$-$\phi$ plane \\
        $\Delta R(\gamma_2, j_2)$ & Distance between sub-leading photon and sub-leading jet in $\eta$-$\phi$ plane \\
        $\Delta R(j_1, j_2)$ & Distance between leading jet and sub-leading jet in $\eta$-$\phi$ plane \\
        $\cos(\theta_{\gamma_1})$ & Cosine of the polar angle $\theta$ of leading photon \\
        $\cos(\theta_{\gamma_2})$ & Cosine of the polar angle $\theta$ sub-leading photon\\
        $\cos(\theta_{j_1})$ & Cosine of the polar angle $\theta$ leading jet\\
        $\cos(\theta_{j_1})$ & Cosine of the polar angle $\theta$ sub-leading jet\\
        $\cos(\theta_{\gamma_1\gamma_2})$ & The cosine of the angle $\theta$ between the leading photon and the sub-leading photon\\
        $\cos(\theta_{j_1j_2})$ & The cosine of the angle $\theta$ between the leading jet and the sub-leading jet \\
        $m_{j_1 j_2}$ & Invariant mass of reconstructed two jet system \\
        $\gamma\gamma$-cent & The centrality of the reconstructed two-photon system relative to the two highest-$p_T$ jets\\
        \hline
        \hline
        \bottomrule
    \end{tabular}
\end{table}

The first row in Fig.~\ref{BDT_output} shows the signal efficiency as a function of background rejection, also known as the Receiver Operating Characteristic (ROC) curve, while second row corresponds to the BDT classifier response. Consequently, the left column of Fig.~\ref{BDT_output} corresponds to the ROC curve and BDT response for the signal $f_{T8}/\Lambda^{4}=0.7$ TeV$^{-4}$ and all relevant backgrounds for trained data, while the right column is for the signal $f_{T8}/\Lambda^{4} = 0.07$ TeV$^{-4}$. It can be clearly seen in the Fig.~\ref{BDT_output}, the BDT classifier demonstrates a good performance in discriminating the signal from backgrounds without being prone to (falling into) the undesirable states of under-fitting or over-fitting. In order to determine an optimal cut on reconstructed BDT response distributions for each aQGCs, we consider the point where statistical significance is maximum in the region where signal efficiency is in the range between 70\% and 80\%. Each signal process, corresponding to a different aQGC value, is individually trained in the BDT classifier using the signal region. A unique optimal BDT cut score is determined for each signal and the relevant SM background processes. 

\begin{figure}[h!]
\includegraphics[scale=0.408]{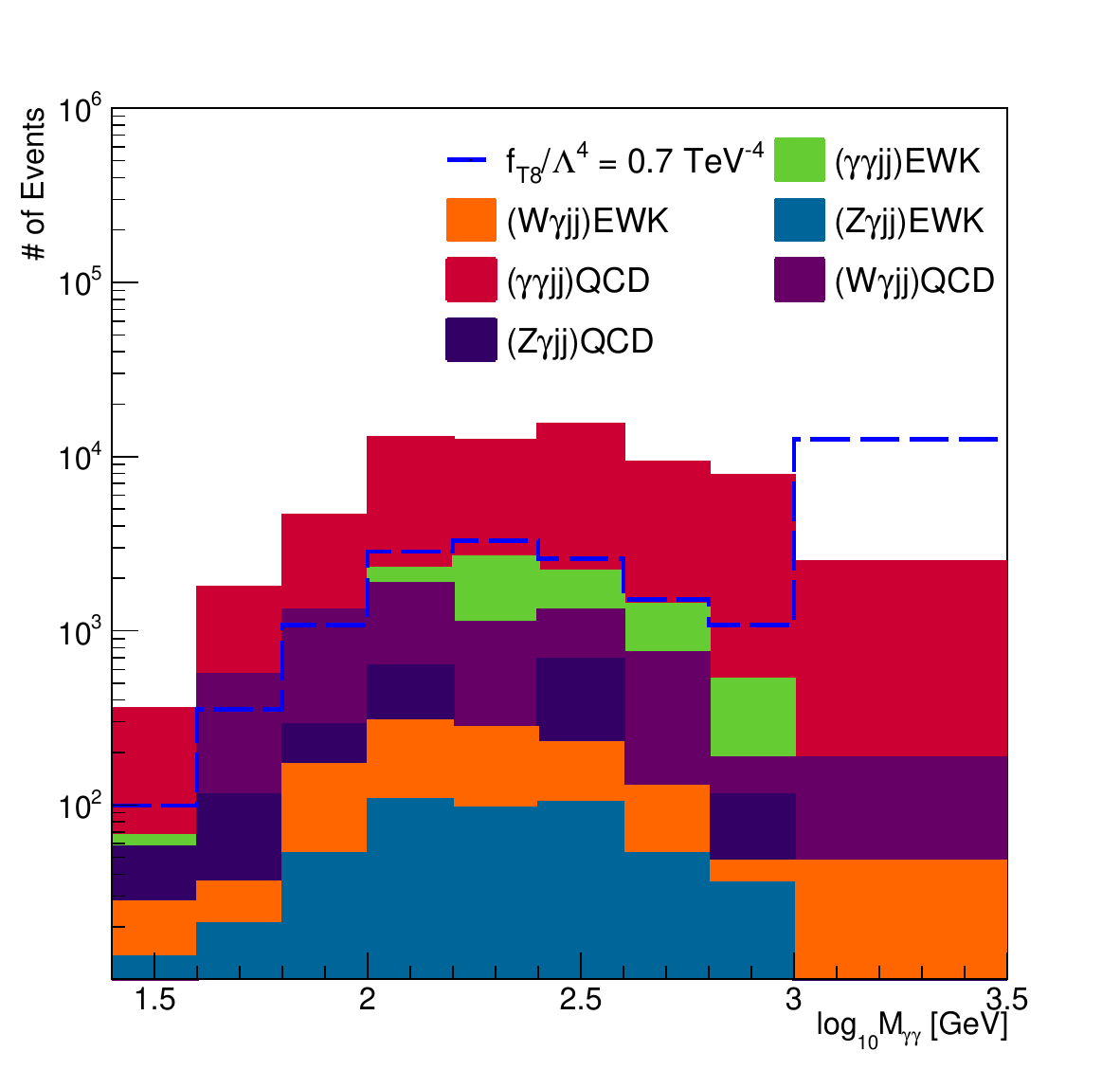}
\includegraphics[scale=0.408]{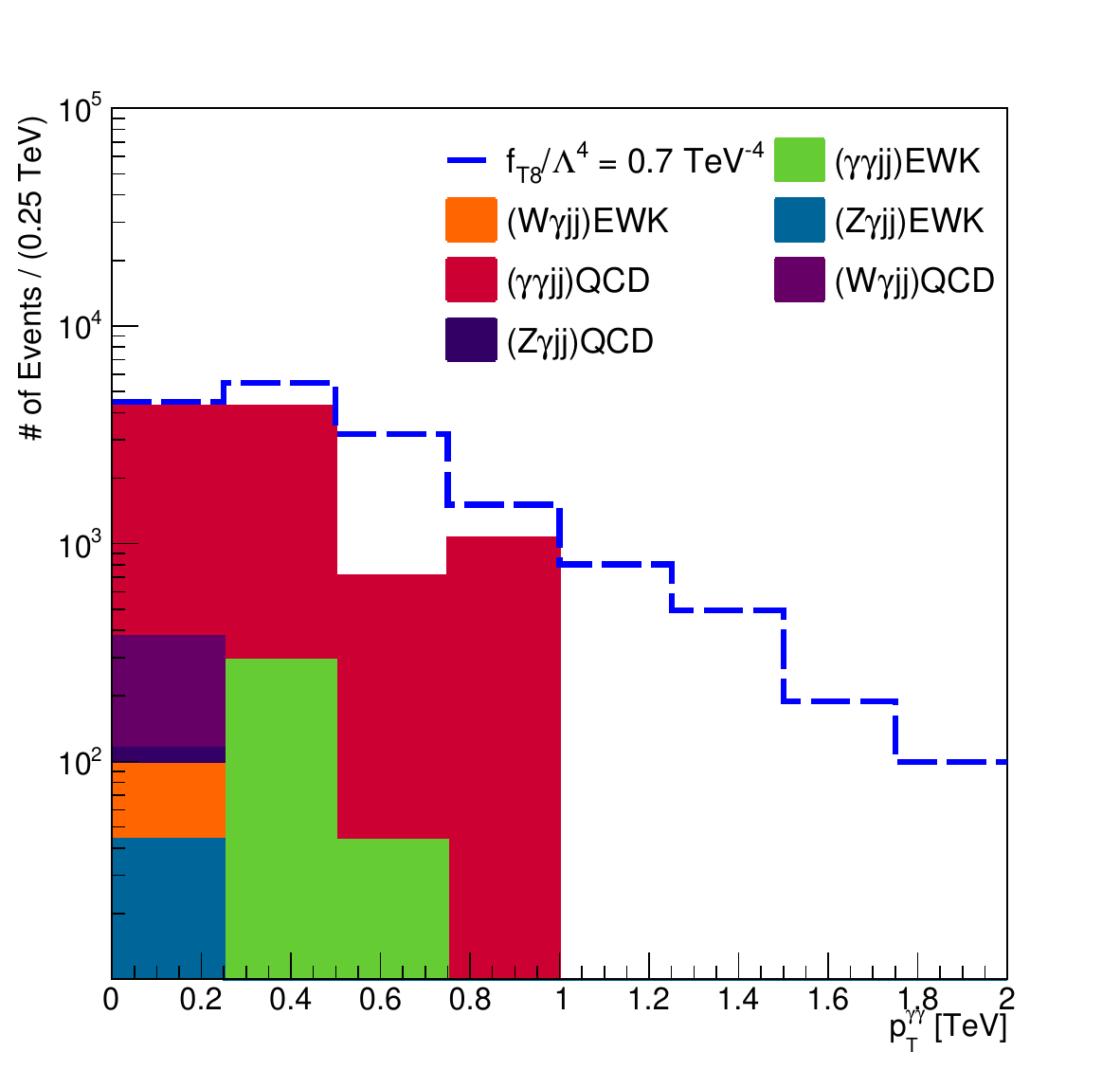}
\includegraphics[scale=0.408]{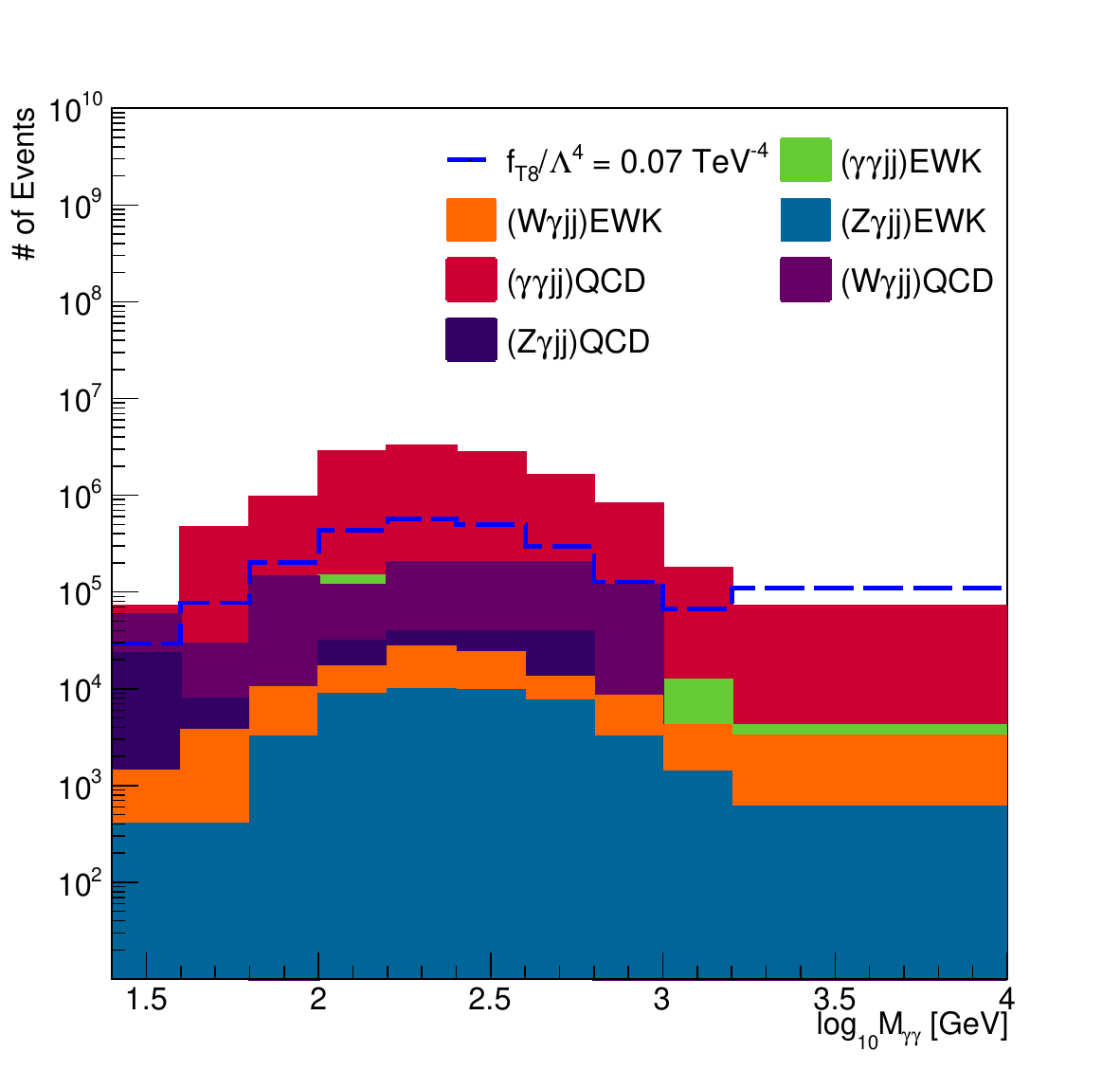}
\includegraphics[scale=0.408]{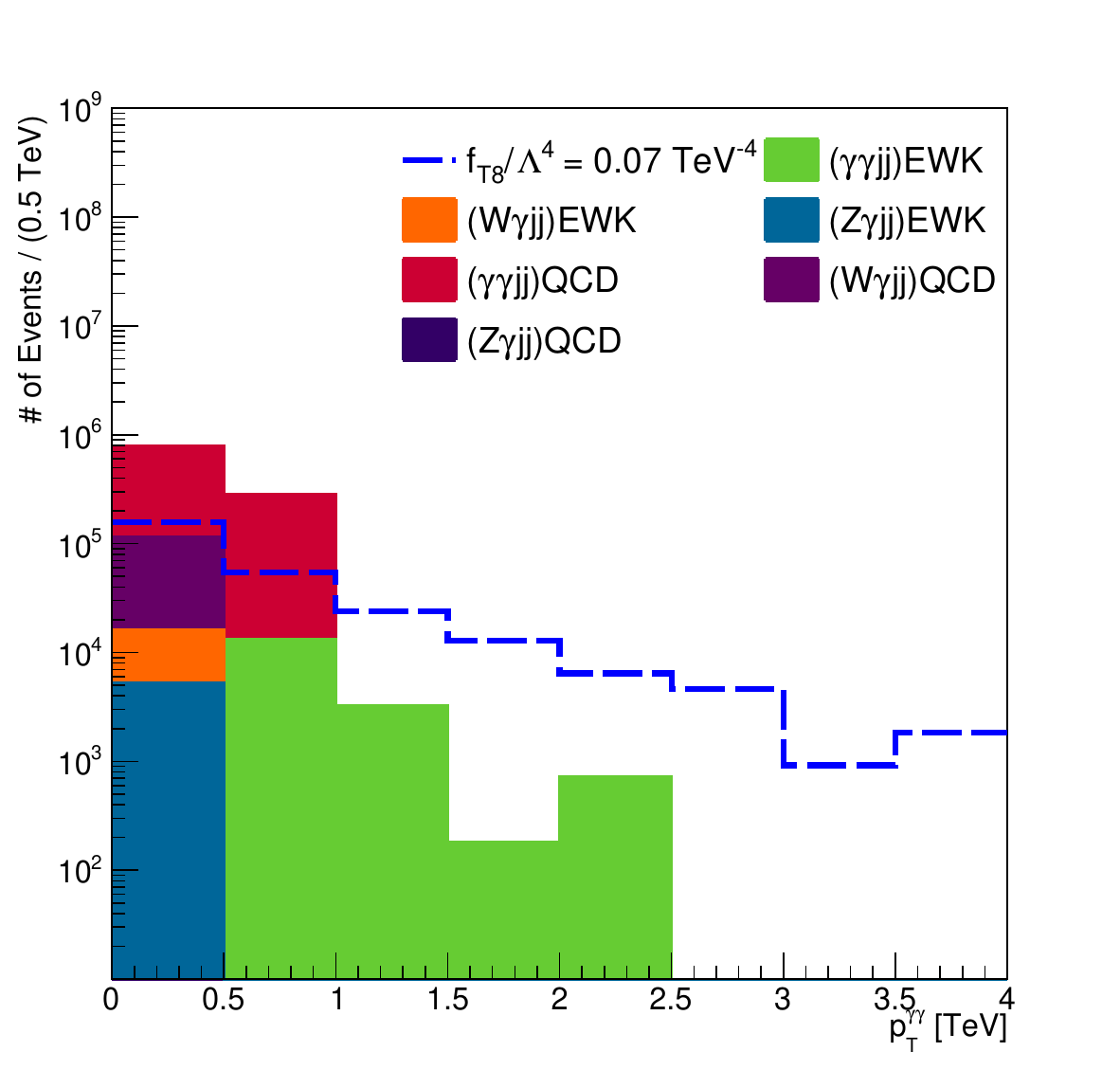}
\caption{Normalized distributions of the invariant mass $(M_{\gamma\gamma}$) and the transverse momentum ($p_T^{\gamma\gamma}$) of the two-photon system for the process $pp \rightarrow \gamma\gamma jj$ at HL-LHC (at the top) and FCC-hh (at the bottom). \label{fig:m_aa-pt_aa}}
\end{figure}

The Table~\ref{noe_hl_lhc} presents the normalized number of events for the signal process \((\gamma\gamma jj){\text{EWK}}\) with an effective coupling parameter of \(f_{T8}/\Lambda^4 = 0.7\) TeV\(^{-4}\), alongside the relevant background processes, after the sequential application of pre-selection, kinematic, reconstructed, and Boosted Decision Tree (BDT) cuts used in Table~\ref{tab_event_selection}. The event numbers are normalized to their respective cross-sections multiplied by the integrated luminosity \(\mathcal{L}_{\text{int}} = 3\) ab\(^{-1}\) at the High Luminosity LHC (HL-LHC) and divided by the total number of generated events.  The pre-selection stage retains a large fraction of events, with the signal yielding approximately \(1.57 \times 10^6\) events, while the dominant background contribution originates from the QCD-induced \((\gamma\gamma jj){\text{QCD}}\) and \((W\gamma jj){\text{QCD}}\) processes, reaching \(1.48 \times 10^8\) and \(1.09 \times 10^8\) events, respectively. The application of kinematic cuts significantly reduces the number of events across all processes, with the signal dropping to \(6.58 \times 10^4\) events, reflecting an efficiency of approximately \(4.2\%\). The background processes experience substantial reductions, with \((\gamma\gamma jj){\text{QCD}}\) decreasing by nearly two orders of magnitude. Reconstructed cuts further enhance the signal-to-background discrimination, reducing the signal to \(3.43 \times 10^4\) events, corresponding to an efficiency of approximately \(52.1\%\) relative to the kinematic stage. The background processes experience even stronger suppression, particularly the QCD-induced \((\gamma\gamma jj){\text{QCD}}\) background, which is reduced to \(4.10 \times 10^5\) events. The final BDT cut significantly improves the purity of the signal by effectively suppressing background contributions. The signal is retained at \(2.55 \times 10^4\) events, while the dominant \((\gamma\gamma jj){\text{QCD}}\) background is reduced to \(6.69 \times 10^4\), corresponding to an efficiency of approximately \(16.3\%\) relative to its reconstructed stage value. Other backgrounds, such as \((W\gamma jj){\text{QCD}}\) and \((Z\gamma jj){\text{QCD}}\), are similarly reduced to \(7.67 \times 10^3\) and \(2.18 \times 10^3\) events, respectively.  
\begin{table}
    \centering
\caption{The number of events for the signal ($f_{T8}/\Lambda^4 = 0.7$ TeV$^{-4}$), as well as all relevant background processes, after applying pre-selection, kinematic, reconstructed  and the optimum BDT cuts. 
The numbers are normalized to the cross-section of each process times the integrated luminosity $\mathcal{L}_{int} = 3$ ab$^{-1}$ for the HL-LHC divided by the total number of generated events for each process, respectively.
\label{noe_hl_lhc}}
\begin{ruledtabular}
\begin{tabular}{lccccccc}
&Processes  & Pre-Selection & Kinematic &Reconstructed & BDT & \textbf{\(\sigma \, [\mathrm{pb}]\)} \\ \hline
Signal &($\gamma \gamma jj$)\textbf{EWK}   & 1.571$\times10^6$ & 6.578$\times10^4$ & 3.425$\times10^4$&2.545$\times10^4$ & 8.191$\times10^{-1}$ \\
\hline
&($\gamma \gamma jj$) \textbf{EWK}   & 1.441$\times10^6$ & 3.679$\times10^4$ & 2.076$\times10^4$& 1.063$\times10^4$ & 7.672$\times10^{-1}$ \\
&($\gamma \gamma jj$) \textbf{QCD}   & 1.477$\times10^8$   & 2.769$\times10^6$ &4.103$\times10^5$& 6.690$\times10^4$ & 1.186$\times10^2$ \\
Backgrounds&($W \gamma jj$) \textbf{EWK}    & 2.349$\times10^6$ & 6.776$\times10^3$ &3.187$\times10^3$& 1.289$\times10^3$ & 1.341$\times10^0$ \\
&($W \gamma jj$) \textbf{QCD} & 1.086$\times10^8$ & 1.762$\times10^5$ &3.965$\times10^4$& 7.669$\times10^3$ & 6.545$\times10^1$ \\
&($Z \gamma jj$) \textbf{EWK}  & 9.908$\times10^5$  & 2.160$\times10^3$ &1.061$\times10^3$& 4.936$\times10^2$ & 5.032$\times10^{-1}$ \\
&($Z \gamma jj$) \textbf{QCD}  & 3.391$\times10^7$  & 3.753$\times10^4$ &8.620$\times10^3$& 2.184$\times10^3$ & 1.916$\times10^1$ \\
\end{tabular}
\end{ruledtabular}
\end{table}

Overall, the sequential application of selection cuts significantly enhances the signal-to-background ratio, demonstrating the effectiveness of the employed kinematic and machine learning-based strategies. Quantitatively, 1.6\% of the pre-selected events remains for further analysis for signal $f_{T8}/\Lambda^4 = 0.7$ TeV$^{-4}$  at HL-LHC. On the other hand, the efficiency of cuts for SM backgrounds are 0.7\% and 0.045\% for EWK and QCD  $\gamma \gamma jj$, 0.055\% and 0.007\% for EWK and QCD  $W \gamma jj$, 0.04\% and 0.064\% for EWK and QCD  $Z \gamma jj$.

Similarly, Table~\ref{noe_fcchh} demonstrates the effects of the cut flow for FCC-hh at an integrated luminosity of \(\mathcal{L}_{\text{int}} = 30\) ab\(^{-1}\) in which the effective coupling parameter set to \(f_{T8}/\Lambda^4 = 0.07\) TeV\(^{-4}\) for the signal and relevant SM background processes. As expected, due to the significantly higher energy and luminosity of the FCC-hh, the overall number of events is substantially larger compared to the HL-LHC scenario.  Similar overall cut effects can be seen in this table. Comparing both collider scenarios, the sequential application of selection cuts plays a crucial role in optimizing event selection and retains a substantial fraction of the signal while suppressing dominant QCD-induced backgrounds.

The optimal BDT cut score along with the cuts in Table \ref{tab_event_selection} is applied to the signal with \( f_{T8}/\Lambda^{4} = 0.7 \) TeV\(^{-4}\) and \( f_{T8}/\Lambda^{4} = 0.07 \) TeV\(^{-4}\), as well as the relevant background processes, to obtain the invariant mass (\( M_{\gamma\gamma} \)) and transverse momentum (\( p_T^{\gamma\gamma} \)) distributions of the two-photon system, presented HL-LHC at the top and FCC-hh at the bottom in Fig.~\ref{fig:m_aa-pt_aa}. The invariant mass distribution (\( M_{\gamma\gamma} \)) reveals a distinct separation between the electroweak signal (dashed blue line) and background contributions. The signal exhibits a pronounced presence at higher mass values, aligning with the expected kinematic behavior of Beyond Standard Model effects. In contrast, QCD-induced background processes, such as \((\gamma\gamma jj){\text{QCD}}\) and \((W\gamma jj){\text{QCD}}\), peak at lower values and diminish significantly in the high-mass region. Applying the \( M_{\gamma\gamma} > 2.8 \) TeV selection cut significantly alters the transverse momentum (\( p_T^{\gamma\gamma} \)) distributions. Since \( p_T^{\gamma\gamma} \) is correlated with \( M_{\gamma\gamma} \), the high-mass requirement naturally shifts the \( p_T^{\gamma\gamma} \) spectrum towards larger values. The signal events remain dominant in the high-\( p_T \) region, while QCD backgrounds are substantially suppressed. The electroweak backgrounds (\((\gamma\gamma jj){\text{EWK}}\), \((W\gamma jj){\text{EWK}}\), and \((Z\gamma jj){\text{EWK}}\)) are less affected by the mass selection, maintaining a relatively stable contribution compared to the QCD-induced backgrounds.

Finally, \( p_T^{\gamma\gamma} \) distributions obtained after applying a selection cut of \( M_{\gamma\gamma} > 2.8 \) TeV is used to evaluate the sensitivities of aQGC.

\begin{table}
    \centering
\caption{The number of events for the signal ($f_{T8}/\Lambda^4 = 0.07$ TeV$^{-4}$), as well as all relevant background processes, after applying pre-selection, kinematic cuts, and the optimum BDT cut. The numbers are normalized to the cross-section of each process times the integrated luminosity $\mathcal{L}_{int} = 30$ ab$^{-1}$ for the FCC-hh divided by the total number of generated events for each process, respectively.
\label{noe_fcchh}}
\begin{ruledtabular}
\begin{tabular}{lccccccc}
&Processes  & Pre-Selection & Kinematic &Reconstructed & BDT & \textbf{\(\sigma \, [\mathrm{pb}]\)} \\ \hline
Signal &($\gamma \gamma jj$)\textbf{EWK}   & 3.282$\times10^8$ &  6.572$\times10^6$ &3.254$\times10^6$&2.419$\times10^6$ &1.583$\times10^1$ \\
\hline
&($\gamma \gamma jj$) \textbf{EWK}   & 1.448$\times10^8$ & 2.333$\times10^6$ & 1.159$\times10^6$& 7.910$\times10^5$ & 5.643$\times10^0$ \\
&($\gamma \gamma jj$) \textbf{QCD}   & 2.004$\times10^{10}$   & 2.975$\times10^8$ &4.489$\times10^7$& 1.302$\times10^7$ & 1.186$\times10^3$\\
Backgrounds&($W \gamma jj$) \textbf{EWK}    & 2.003$\times10^8$ & 3.921$\times10^5$ & 1.624$\times10^5$&1.152$\times10^5$  & 1.491$\times10^1$ \\
&($W \gamma jj$) \textbf{QCD} & 1.390$\times10^{10}$ & 1.825$\times10^7$ &4.126$\times10^6$& 1.075$\times10^6$ & 9.689$\times10^2$ \\
&($Z \gamma jj$) \textbf{EWK}  & 1.113$\times10^8$  & 1.452$\times10^5$ &6.446$\times10^4$& 4.569$\times10^4$ & 6.583$\times10^0$ \\
&($Z \gamma jj$) \textbf{QCD}  & 4.151$\times10^9$  & 2.559$\times10^6$ & 5.289$\times10^5$& 1.788$\times10^5$ & 2.592$\times10^2$ \\
\end{tabular}
\end{ruledtabular}
\end{table}

\section{Limits on Anomalous Quartic Gauge Couplings}\label{secIV}

\begin{figure}[httb!]
\includegraphics[scale=0.27]
{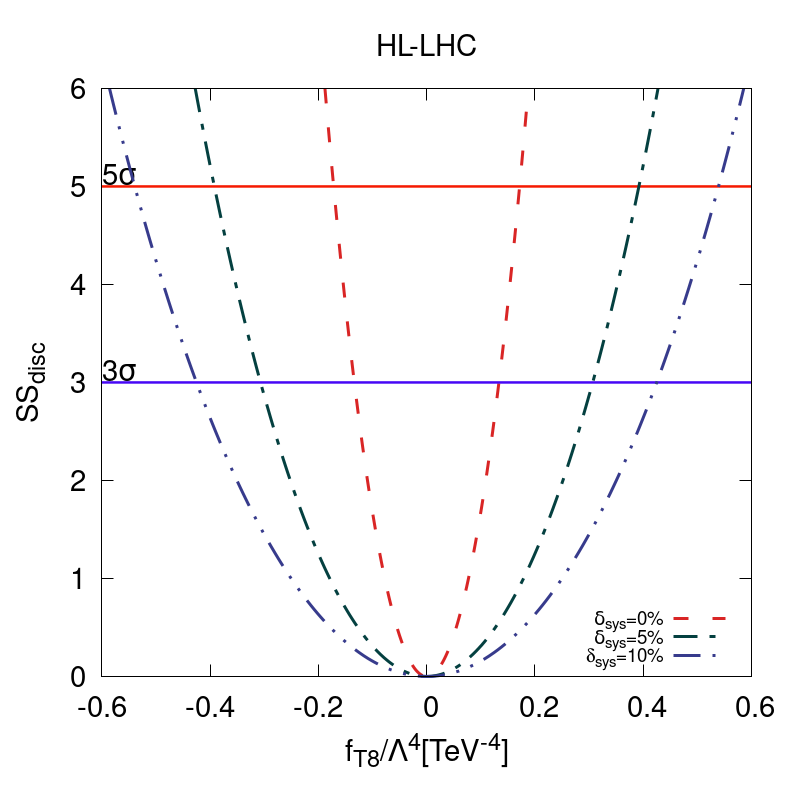} 
\includegraphics[scale=0.27]
{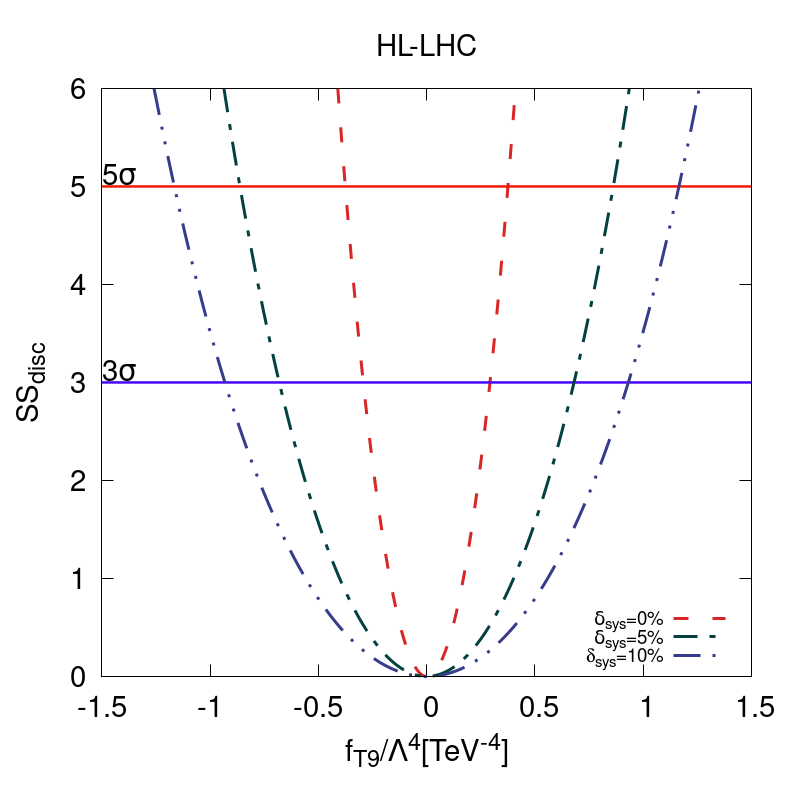} 
\includegraphics[scale=0.27]
{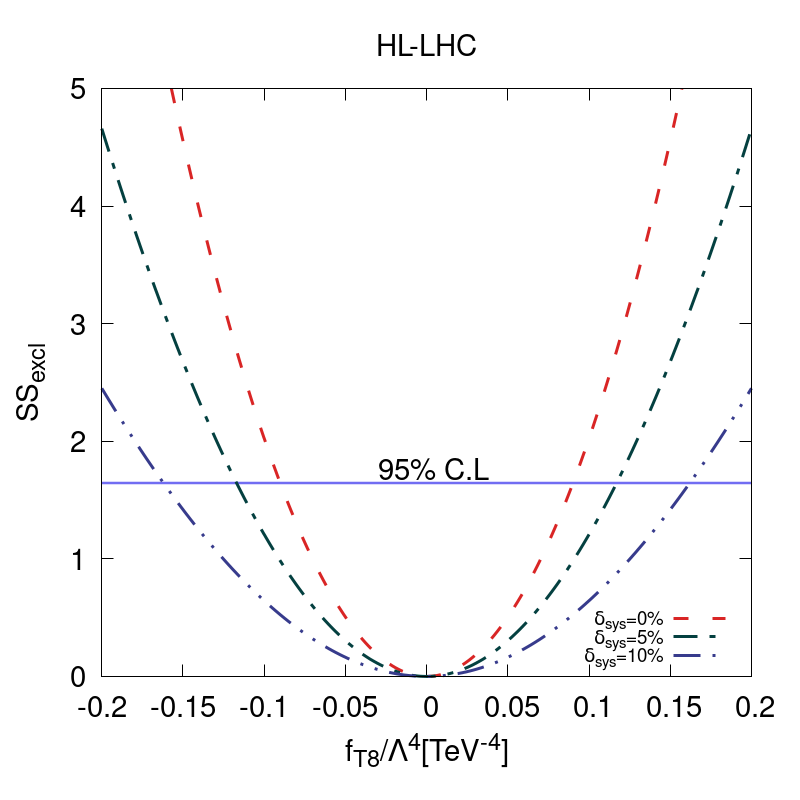} 
\includegraphics[scale=0.27]
{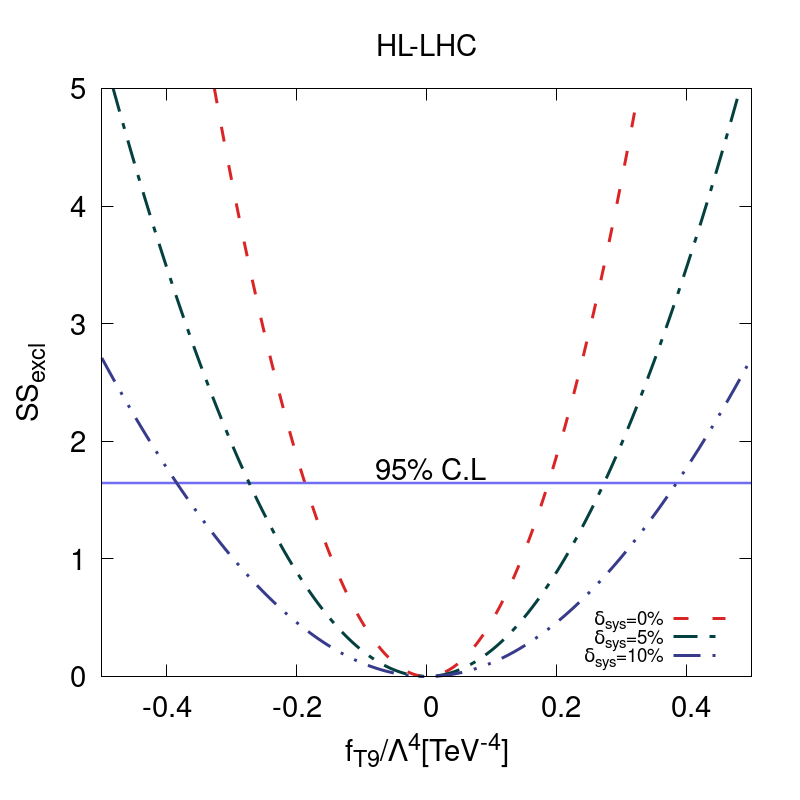} 
\caption{The $\mathcal{SS}_{disc}$ and $\mathcal{SS}_{excl}$ as a function of  $f_{T8}/\Lambda^4$ and $f_{T9}/\Lambda^4$  couplings for $pp \rightarrow \gamma\gamma jj$ process with and without systematic uncertainty at HL-LHC ($\sqrt{s}$=14 TeV). \label{fig:ss_hl}}
\end{figure}

\begin{figure}[httb!]
\includegraphics[scale=0.27]{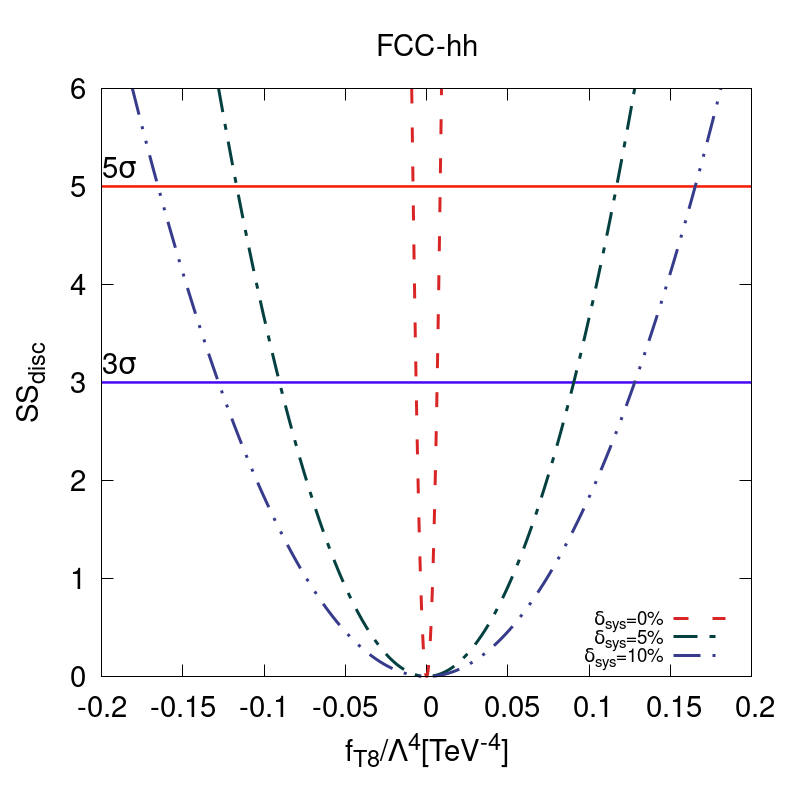}
\includegraphics[scale=0.27]{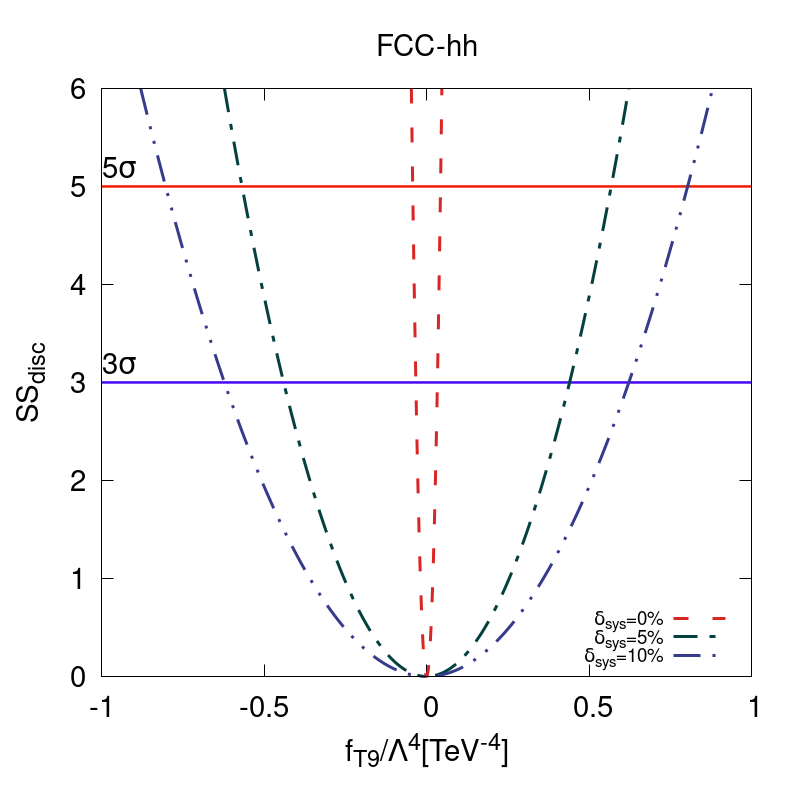}
\includegraphics[scale=0.27]{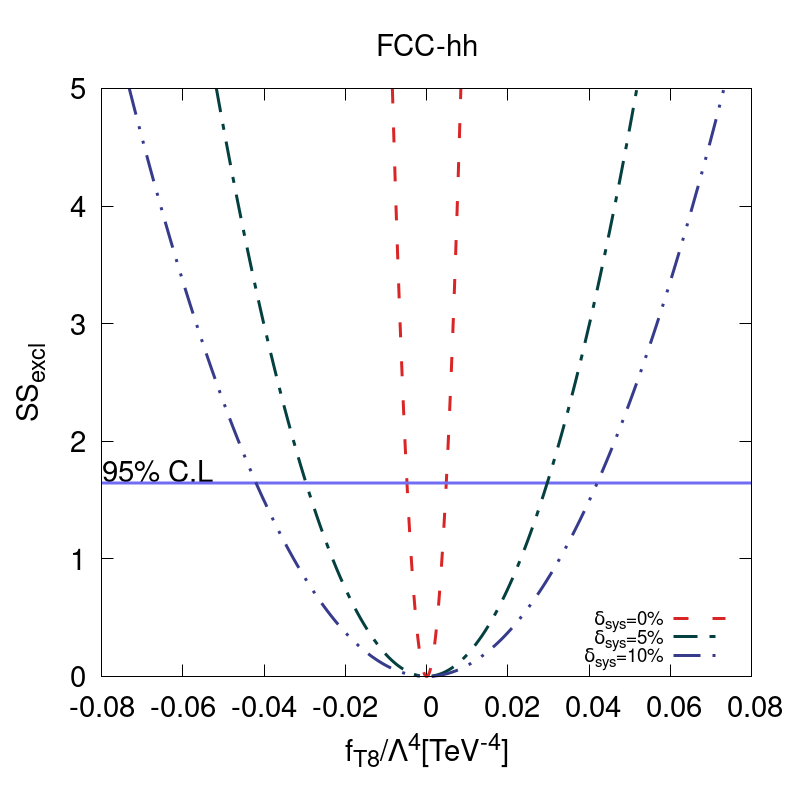}
\includegraphics[scale=0.27]{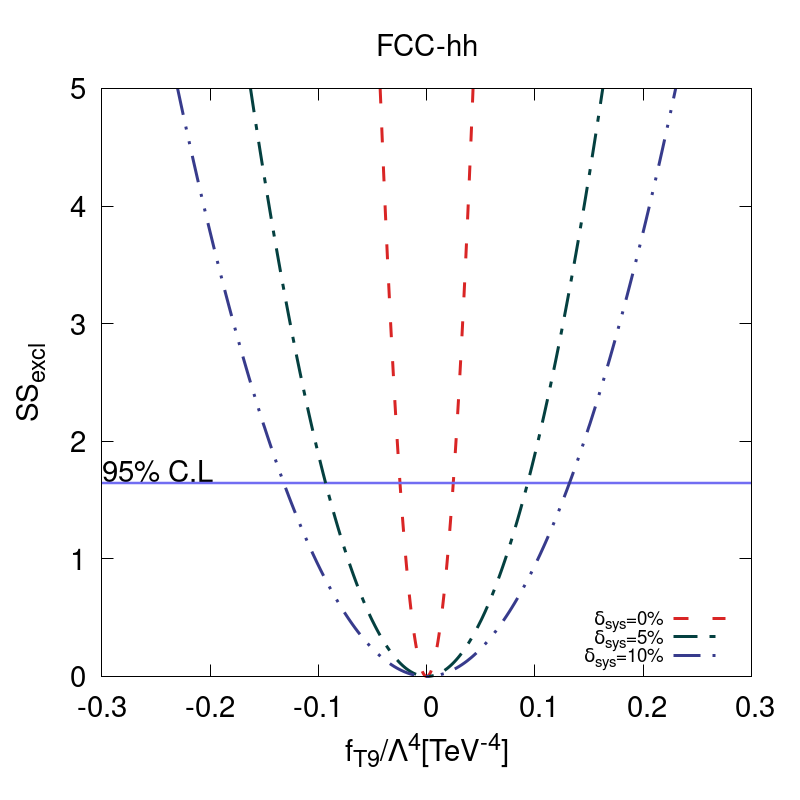}
\caption{The $\mathcal{SS}_{disc}$ and $\mathcal{SS}_{excl}$ as a function of  $f_{T8}/\Lambda^4$ and $f_{T9}/\Lambda^4$ couplings for $pp \rightarrow \gamma\gamma jj$ process with and without systematic uncertainty at FCC-hh ($\sqrt{s}$= 100 TeV). \label{fig:ss_fcc}}
\end{figure}

In this section, we report the discovery and exclusion reach of the aQGCs, $f_{T8}/\Lambda^4$ and $f_{T9}/\Lambda^4$, for the process $pp \rightarrow \gamma\gamma jj$ at HL-LHC and FCC-hh. Expressions to obtain median expected significance for discovery ($\mathcal{SS}_{disc}$) and exclusion ($\mathcal{SS}_{excl}$) of a new phenomena including systematic uncertainties in high energy physics \cite{Cowan:2010js,Kumar:2015tna}

\begin{equation}
\mathcal{SS}_{\text{disc}} = \sqrt{2 \left[ (S+B) \ln \left (\frac{(S+B)(1 + \delta^{2}B)}{B + \delta^{2} B (S+B)} \right) -\frac{1}{\delta^{2}} \ln \left(1 + \delta^{2} \frac{S}{1 + \delta^{2}B}\right)\right ] }
\end{equation}
\begin{equation}
\mathcal{SS}_{\text{excl}} = \sqrt{2 \left[S - B \ln \left( \frac{B+S+x}{2B}\right) -\frac{1}{\delta^{2}} \ln \left ( \frac{B-S+x}{2B}\right) \right] - (B+S-x)(1 + \frac{1}{\delta^{2}B})}
\end{equation}
where $x = \sqrt{(S + B)^{2} - 4\delta^{2}SB^{2}/(1 + \delta^{2}B)}$. $S$ and $B$ are the number of signal and total SM background events, obtained by the normalized transverse momentum ($p_{T}^{\gamma \gamma}$) distribution of the two-photon system (right panel in Fig.~\ref{fig:m_aa-pt_aa}) and $\delta$ is the systematic uncertainty. In the limit $\delta \rightarrow$ 0, these expressions can be simplified as
\begin{equation}
\mathcal{SS}_{\text{disc}} = \sqrt{2 \left[(S + B) \ln \left ( 1 + S/B \right) - S \right] }
\end{equation}
\begin{equation}
\mathcal{SS}_{\text{excl}} = \sqrt{2 \left[S - B \ln \left ( 1 + S/B \right) \right]}
\end{equation}
Region with $\mathcal{SS}_{disc}$ $\geq$ 3 (5) $\sigma$ are classified as 
discoverable regions, while region with a $\mathcal{SS}_{excl}$ $\leq$1.645 are considered regions that can be excluded at a 95\% C.L. Figs.~\ref{fig:ss_hl} and \ref{fig:ss_fcc} illustrate the statistical significance for both $SS_{disc}$  and $SS_{excl}$ as functions of the anomalous couplings $f_{T8}/\Lambda^4$ (on the left) and $f_{T9}/\Lambda^4$ (on the right) couplings for the process $pp \to \gamma\gamma jj$, presented for two collider scenarios: the HL-LHC with a centre-of-mass energy of $\sqrt{s} = 14$ TeV and the FCC-hh at $\sqrt{s} = 100$ TeV , respectively. When assessing the realistic physics potential of future colliders for the $pp \rightarrow \gamma\gamma jj$ process, it is crucial to account for the impact of systematic uncertainties on the limits. These uncertainties primarily stem from cross-section measurements of the $pp \rightarrow \gamma\gamma jj$ process based on leading-order (LO) or next-to-leading-order (NLO) predictions, as well as higher-order electroweak corrections, uncertainties in the integrated luminosity, and misidentifications of electrons and jets as photons. In this study, we focus on LO predictions and do not explore the effects or validity of these higher-order corrections on both the signal and the Standard Model background processes.
Each plot displays the impact of different levels of systematic uncertainty ($\delta_{syst} = 0\%$, $5\%$, and $10\%$) on the significance curves. The upper panels in both figures correspond to discovery potential, where horizontal lines indicate the $3\sigma$ and $5\sigma$ thresholds, while the lower panels show exclusion limits with the $95\%$ confidence level (C.L.) marked. The results highlight how increasing systematic uncertainties affect the sensitivity to deviations from the Standard Model predictions. 
The limits on the anomalous quartic gauge couplings $f_{T8}/\Lambda^4$ and $f_{T9}/\Lambda^4$ have been determined by the intersection points of the significance curves with the $3\sigma$, $5\sigma$, and $95\%$ C.L. reference lines  in Figs.~\ref{fig:ss_hl} and \ref{fig:ss_fcc}. The corresponding numerical values are summarized in Tables~\ref{limits_hllhc} and \ref{limits_fcchh} for the HL-LHC and FCC-hh scenarios, respectively. Table~\ref{limits_hllhc} presents the obtained limits at HL-LHC for an integrated luminosity of $\mathcal{L}_{\text{int}} = 3\, \text{ab}^{-1}$. The results indicate that the sensitivity to the anomalous couplings degrades as the level of systematic uncertainty $\delta_{\text{sys}}$ increases from $0\%$ to $10\%$. At the $3\sigma$ significance level, the constraints on $f_{T8}/\Lambda^4$ vary from $[-1.06, 1.06] \times 10^{-1}$~TeV$^{-4}$ in the absence of systematic uncertainty to $[-4.43, 4.43] \times 10^{-1}$~TeV$^{-4}$ when $\delta_{\text{sys}} = 10\%$. A similar trend is observed for $f_{T9}/\Lambda^4$, where the limits become significantly weaker with increasing systematic uncertainty. The constraints
on $f_{T9}/\Lambda^4$ coupling are $[-2.11, 2.11] \times 10^{-1}$~TeV$^{-4}$, $[-5.99, 5.99] \times 10^{-1}$~TeV$^{-4}$ and $[-8.53, 8.53] \times 10^{-1}$~TeV$^{-4}$ without and with systematic uncertainties $\delta_{\text{sys}} = 5\%$ $\delta_{\text{sys}} = 10\%$,  respectively. Table~\ref{limits_fcchh} provides the limits corresponding to $\mathcal{L}_{\text{int}} = 30\, \text{ab}^{-1}$ for the FCC-hh collider. Due to the higher centre-of-mass energy and luminosity, the sensitivity to the anomalous couplings is greatly enhanced compared to HL-LHC. At $3\sigma$ significance, the bounds on $f_{T8}/\Lambda^4$ ($f_{T9}/\Lambda^4$) coupling improve to $[-6.48, 6.48] \times 10^{-3}$~TeV$^{-4}$ $([-3.32, 3.32] \times 10^{-2}$~TeV$^{-4})$ for $\delta_{\text{sys}} = 0\%$. Even with the inclusion of a $10\%$ systematic uncertainty, the obtained limits remain significantly stronger than those at HL-LHC. This demonstrates the superior potential of FCC-hh in probing deviations from the Standard Model. In both collider scenarios, the exclusion limits at $95\%$ C.L. exhibit a similar dependence on systematic uncertainties. As expected, FCC-hh provides the most stringent constraints on $f_{T8}/\Lambda^4$ and $f_{T9}/\Lambda^4$, reinforcing its capability to explore new physics effects with higher precision.
\begin{table}[httb!]
\centering
\caption{The limits on anomalous quartic gauge couplings in units of [TeV]$^{-4}$ at $3\sigma$, $5\sigma$ and 95 \%\ \text{C.L}, considering $\delta_{sys}$ = 0, 5\% and 10\% systematic uncertainties for $\mathcal{L}_{int} = 3$ ab$^{-1}$ at HL-LHC.\label{limits_hllhc}}
\begin{ruledtabular}
\begin{tabular}{c c c c c}
Coefficient  & $\delta_{\text{sys}}$ & $3\sigma$ & $5\sigma$ & $95 \% \, \text{C.L.}$ \\ \hline
\multirow{3}{*}{$f_{T8}/\Lambda^{4}$} & 0\% & [-1.06; 1.06]$\times 10^{-1}$ & [-1.36; 1.36]$\times 10^{-1}$ & [-9.02; 9.02]$\times 10^{-2}$ \\ 
 & 5\% & [-3.12; 3.12]$\times 10^{-1}$ & [-3.96; 3.96]$\times 10^{-1}$ & [-1.17; 1.17]$\times 10^{-1}$ \\ 
 & 10\% & [-4.43; 4.43]$\times 10^{-1}$ & [-5.58; 5.58]$\times 10^{-1}$ & [-1.62; 1.62]$\times 10^{-1}$ \\ \hline
\multirow{3}{*}{$f_{T9}/\Lambda^{4}$} & 0\% & [-2.11; 2.11 ]$\times 10^{-1}$ & [-2.73; 2.73 ]$\times 10^{-1}$ & [-1.87 ; 1.87]$\times 10^{-1}$ \\ 
 & 5\% & [-5.99; 5.99]$\times 10^{-1}$ & [-7.67; 7.67]$\times 10^{-1}$ & [-2.73; 2.73]$\times 10^{-1}$ \\ 
 & 10\% & [-8.53; 8.53]$\times 10^{-1}$ & [-1.08; 1.08]$\times 10^{0}$ & [-3.85; 3.85]$\times 10^{-1}$
\end{tabular}
\end{ruledtabular}
\end{table}

\begin{table}[httb!]
\centering
\caption{The limits on anomalous quartic gauge couplings in units of [TeV]$^{-4}$ at $3\sigma$, $5\sigma$ and 95 \%\ \text{C.L}, considering $\delta_{sys}$ = 0, 5\% and 10\% systematic uncertainties for $\mathcal{L}_{int} = 30$ ab$^{-1}$ at FCC-hh.\label{limits_fcchh}}
\begin{ruledtabular}
\begin{tabular}{c c c c c}
Coefficient  & $\delta_{\text{sys}}$ & $3\sigma$ & $5\sigma$ & $95 \% \, \text{C.L.}$ \\ \hline
\multirow{3}{*}{$f_{T8}/\Lambda^{4}$} & 0\% & [-6.48; 6.48]$\times 10^{-3}$ & [-8.37; 8.37]$\times 10^{-3}$ & [-4.84; 4.84]$\times 10^{-3}$ \\ 
 & 5\% & [-9.06; 9.06]$\times 10^{-2}$ & [-1.17; 1.17]$\times 10^{-1}$ & [-2.97; 2.97]$\times 10^{-2}$ \\ 
 & 10\% & [-1.28; 1.28]$\times 10^{-1}$ & [-1.65; 1.65]$\times 10^{-1}$ & [-4.20; 4.20]$\times 10^{-2}$ \\ \hline
\multirow{3}{*}{$f_{T9}/\Lambda^{4}$} & 0\% & [-3.32; 3.32]$\times 10^{-2}$ & [-4.29; 4.29]$\times 10^{-2}$ & [-2.46; 2.46]$\times 10^{-2}$ \\ 
 & 5\% & [-4.40; 4.40]$\times 10^{-1}$ & [-5.68; 5.68]$\times 10^{-1}$ & [-9.33; 9.33]$\times 10^{-2}$ \\ 
 & 10\% & [-6.22; 6.22]$\times 10^{-1}$ & [-8.04; 8.04]$\times 10^{-1}$ & [-1.32; 1.32]$\times 10^{-1}$
\end{tabular}
\end{ruledtabular}
\end{table}
The best experimental limits at 95 \% C.L. reported by the ATLAS collaboration \cite{ATLAS:2022nru}, under the assumption that unitarity is preserved and setting all other aQGCs to zero except one, are $[-5.2,5.2] \times 10^{-1}$~TeV$^{-4}$ for $f_{T8}/\Lambda^{4}$ and $[-7.9,7.9] \times 10^{-1}$~TeV$^{-4}$ for $f_{T9}/\Lambda^{4}$. If unitarity is not enforced, the reported limits improve to $[-5.9,5.9] \times 10^{-2}$~TeV$^{-4}$ and $[-1.3,1.3] \times 10^{-1}$~TeV$^{-4}$ for $f_{T8}/\Lambda^{4}$ and $f_{T9}/\Lambda^{4}$, respectively. In comparison with experimental limits, the projected sensitivity obtained from our analysis at HL-LHC ($\sqrt{s} = 14$ TeV, $\mathcal{L}_{\text{int}} = 3$~ab$^{-1}$) and FCC-hh ($\sqrt{s} = 100$ TeV, $\mathcal{L}_{\text{int}} = 30$~ab$^{-1}$) significantly improves upon these experimental results. As seen in Table~\ref{limits_hllhc}, the HL-LHC limits on $f_{T8}/\Lambda^{4}$ reach the order of $[-9.02,9.02] \times 10^{-2}$~TeV$^{-4}$ at a 95\% C.L. without systematic uncertainties, which is already tighter than the current ATLAS bounds with unitarity violation. Similarly, the 95\% C.L. limits for $f_{T9}/\Lambda^{4}$ at HL-LHC, given as $[-1.87,1.87] \times 10^{-1}$~TeV$^{-4}$, also provide a substantial improvement over the existing experimental constraints. The projected limits become even more stringent at FCC-hh as summarized in Table~\ref{limits_fcchh}. The bounds on $f_{T8}/\Lambda^{4}$ are expected to reach $[-4.84,4.84] \times 10^{-3}$~TeV$^{-4}$ at 95\% C.L. significance when systematic uncertainties are neglected. This represents an improvement of nearly two orders of magnitude compared to the best available ATLAS limits. Similarly, for $f_{T9}/\Lambda^{4}$, FCC-hh projections indicate limits at the level of $[-2.46,2.46] \times 10^{-2}$~TeV$^{-4}$, showing a drastic enhancement in sensitivity.

\section{Conclusion}\label{secV}

Investigating quartic vector boson couplings serves not only as a critical test of the Standard Model (SM) but also as a potential probe for new physics phenomena at higher energy scales. These effects can be effectively parameterized using higher-order operators within the framework of Effective Field Theory (EFT). Moreover, future hadron-hadron colliders, with their anticipated advancements in energy reach, luminosity, and detector performance, offer a promising avenue for extending the sensitivity to deviations from SM predictions at unprecedented energy scales.  

In this study, we focus on the Vector Boson Scattering (VBS) process $pp \rightarrow \gamma\gamma jj$ to explore the sensitivity to dimension-8 Anomalous Quartic Gauge Couplings (aQGCs), particularly $f_{T8}/\Lambda^4$ and $f_{T9}/\Lambda^4$. Given that purely electroweak (EW) VBS processes at leading order (LO) are characterized by two outgoing vector bosons and widely separated scattered quark jets with a large invariant mass ($m_{jj}$), they provide a unique experimental signature that facilitates signal isolation from background processes. To ensure a comprehensive analysis, we also consider the dominant background contributions, including electroweak and QCD-induced $\gamma\gamma jj$ production, as well as $W\gamma jj$ and $Z\gamma jj$ processes from both electroweak and QCD interactions. Signal and background events are generated using the MadGraph framework, incorporating aQGC effects, followed by parton showering and hadronization in Pythia. Detector effects are simulated using Delphes with collider-specific parameter tuning.  

To enhance the sensitivity to $f_{T8}/\Lambda^4$ and $f_{T9}/\Lambda^4$, we perform a multivariate analysis using Boosted Decision Trees (BDT) within the TMVA framework. A set of 22 kinematic observables, including transverse momenta and pseudo-rapidities of final-state particles, as well as reconstructed variables such as the invariant mass of the dijet system ($m_{j_1 j_2}$) and the centrality of the diphoton system relative to the two leading jets ($\gamma\gamma$-centrality), is used as input for classification. Furthermore, unitarity-violating effects associated with dimension-8 aQGCs are investigated by applying appropriate energy cutoffs to the invariant mass of the diphoton system.  

Sensitivity projections are derived based on the median expected significance for discovery ($\mathcal{SS}_{disc}$) and exclusion ($\mathcal{SS}_{excl}$), incorporating systematic uncertainties to provide a realistic assessment of the physics reach of future collider experiments. Our results indicate that the FCC-hh imposes more stringent constraints compared to the HL-LHC, reflecting its superior discovery and exclusion potential for probing aQGCs. Notably, even when a 10\% systematic uncertainty is considered, our 95\% C.L. limits remain approximately an order of magnitude more stringent than those recently reported by the ATLAS collaboration \cite{ATLAS:2022nru}, underscoring the enhanced discovery potential of next-generation colliders.

\begin{acknowledgments}
This work was supported by the Scientific and Technological Research Council of Turkey Turkish (TUBITAK), Grant No: 124F116. The numerical calculations reported in this paper were partially performed at TUBITAK ULAKBIM, High Performance and Grid Computing Center (TRUBA resources) and at BAIBU High Energy Physics Laboratory server, received by TUBITAK (Grant No: 114F138)
\end{acknowledgments}

\end{document}